\documentclass[12pt,a4paper]{article}
\usepackage{amsmath, amssymb, amscd, amsthm, amsfonts}
\usepackage{booktabs, diagbox, authblk, orcidlink}
\usepackage{caption}
\usepackage{subcaption}
\usepackage{hyperref}
\usepackage{stackengine}
\usepackage[utf8]{inputenc}
\usepackage{natbib}
\usepackage{float}
\usepackage{authblk}
\usepackage{booktabs}
\usepackage{geometry}

\usepackage{xcolor}
\usepackage{subcaption}
\usepackage{longtable, booktabs}
\usepackage{hyperref}

\title{}

\date{}

\title{A mathematical model of vaccine hesitancy: Analysing the impact of political trends and the interaction across age and education groups in the USA

}
\author{Wajid Ali$^{\orcidlink{0000-0001-5533-1315}1\dagger*}$, Javad Mohamadichamgavi$^{\orcidlink{0000-0001-7208-5996}2\dagger }$, Juan Marcos Ramírez$^{\orcidlink{0000-0003-0000-1073}3}$, Jose Aguilar$^{\orcidlink{0000-0003-4194-6882}3,4}$, Antonio Fernández Anta$^{\orcidlink{0000-0001-6501-2377}5,3}$, Kieran J. Sharkey$^{\orcidlink{0000-0002-7210-92461}1}$}
\affil{
{\footnotesize
$^1$Department of Mathematical Sciences, University of Liverpool, Peach Street, Liverpool, L69 7ZL, England, United Kingdom

$^2$Institute of Applied Mathematics and Mechanics,
University of Warsaw, Banacha 2, 02-097 Warsaw, Poland

$^3$IMDEA Networks Institute, Avenida del Mar Mediterráneo, 22, Leganes, 28918, Madrid, Spain

$^4$Universidad de Los Andes, Mérida, Venezuela

$^5$ IMDEA Software Institute, Campus Montegancedo,
28223, Pozuelo de Alarcón, Madrid, Spain

$^\dagger$ These authors have contributed equally to this work.

$^*$Corresponding author: \texttt{wajid.ali@liverpool.ac.uk}
}}

\date{}
\begin{document}
\maketitle


\begin{abstract}
Vaccination against the SARS-CoV-2 disease has significantly reduced its mortality rate and spread. However, despite its availability, a considerable proportion of the public has either refused or delayed getting vaccinated. This reluctance is known as \textit{vaccine hesitancy}. 
The aim of this paper is to present a mathematical model to investigate how social interaction can impact vaccine hesitancy. 
The model describes the temporal transitions between different vaccination classes of the population (those vaccinated, those who are not yet vaccinated but agree to be vaccinated, and those who refuse).
We apply the model to state and national survey data from the USA to estimate model parameters that quantify the rates at which public opinion on vaccination changes. 
Moreover, we investigate how political trends and demographic factors, such as age and education, impact these parameters. Our results show that state-level political affiliation, age, and educational level shape opinions on vaccination and have a strong influence on the temporal dynamics of attitude changes.
\end{abstract}
\section{Introduction}
The COVID-19 outbreak caused by the SARS-CoV-2 virus started in Wuhan, China, in late 2019~\cite{Zhang2020}. The virus spread quickly worldwide and was classified as a pandemic by the World Health Organization (WHO) in early March 2020~\cite{Cucinotta2020WHOPandemic}. In the United States of America (USA), the virus caused over 100 million confirmed cases of COVID-19 and about 1.1 million deaths by the end of 2022~\cite{johnshopkins2022}. To reduce the spread, policymakers could use a range of non-pharmaceutical interventions (NPIs), including isolation of cases, social distancing and `lockdowns', encouraging the use of face masks, hand hygiene, and travel restrictions, among others~\cite{Asamoah2020, Dobrovolny2020, Haushofer2020, Moore2021,   Pinky2020, Reis2020, Wang2020}.

Despite the effectiveness of NPIs in slowing the spread of disease, vaccination held the promise of an effective and sustainable approach to prevent widespread morbidity and mortality~\cite{Gao2020, Mohamed2022}. As of January 2023, over 670 million COVID-19 vaccines had been administered in the USA, with approximately 78\%  of the population having received at least one dose and 67\%  being fully vaccinated. Early in 2021, when COVID-19 vaccines first became available to adults in the USA, their willingness to get vaccinated varied significantly. Some sought vaccination immediately while others preferred to wait -- an issue known as \textit{vaccine hesitancy}~\cite{Hammershaimb2023,Painter2021,Lindholt2021}. This hesitancy encompasses a range of intentions, from delaying vaccination to outright refusal. Several factors have been shown to influence COVID-19 vaccine hesitancy, including ethnicity, working status, religious beliefs, political views, gender, age, education, and income~\cite{Amin2017,Pierri2022,Troiano2021}. To promote vaccine uptake during public health emergencies, it is crucial to understand the dynamics of vaccination decision-making, the factors driving different vaccination intentions, and the changes in those intentions. From a policy perspective, targeting hesitant but not refusing individuals may be the most effective initial strategy because they might be more easily persuaded to vaccinate. 

Globally, while the overall number of people who have received a COVID-19 vaccine has increased over time, the rates vary significantly between countries. There is also considerable variation at the state level within the USA, with vaccination rates by the end of 2022 ranging from 53\% in Wyoming to 84\% in Massachusetts~\cite{holder2023}. This variation becomes even more pronounced at more localised spatial levels. These disparities in vaccination rates across and within countries have led to investigations of the factors that cause the delay or refusal of vaccines, even when they are available. Thus, numerous studies have been conducted to measure vaccine hesitancy across different vaccination programs, considering various countries' economic, social, and demographic contexts. Early research on COVID-19 vaccine acceptance, including approaches such as game models, cross-sectional surveys, and cognitive valuation surveys, identified multiple factors that influence vaccine acceptance or uptake~\cite{Hadisoemarto2013, Kabir2021, Kabir2023, Palanca-Tan2008, Yeo2018}. More recent studies have shown that vaccine hesitancy is also affected by social behaviours shaped by cultural, social, and political differences across countries in the decision-making process regarding vaccination. It has been suggested that the politicisation of attitudes toward COVID-19 vaccination has contributed to the geographic differences in vaccination coverage within the USA~\cite{ronn2023vaccination}. Areas with higher support for the Democratic candidate in the 2020 presidential election tended to have higher vaccination rates. Voting patterns have also been linked to differences in mobility and attitudes toward public health measures during the pandemic~\cite{Kabir2023}.

Most research on vaccine hesitancy has relied on conventional methods to gather data from individuals or groups through surveys or interviews at one point in time~\cite{Dube2014, Hoy2022, Soares2021,barello2020vaccine,siegler2021trajectory,fedele2021covid,wong2021covid,sherman2021covid,lazarus2023survey,alhassan2021covid,schwarzinger2021covid,nehal2021worldwide,khubchandani2021covid,yasmin2021covid}. For instance, ~\cite{sherman2021covid} used a cross-sectional survey in the UK to investigate factors influencing COVID-19 vaccination intention. They found that beliefs about vaccine safety, the feeling of having enough information to make an informed decision, and general attitudes towards vaccination significantly shaped people’s intentionsSimilarly, ~\cite{khubchandani2021covid} employed an online survey to explore vaccine hesitancy in the USA, revealing its association with sociodemographic factors, political affiliation, and the perceived threat of infection. However, these methods often overlook vaccine hesitancy's dynamic and context-specific nature, as highlighted by ~\cite{wong2021covid}, who emphasised the need for interventions tailored to local populations. Moreover, large-scale, multi-country surveys, such as that conducted by \cite{lazarus2023survey}, have shown that vaccine acceptance varies over time and between regions, underlining the necessity of adaptive strategies to promote vaccination. Studies in Ghana ~\cite{alhassan2021covid} and France ~\cite{schwarzinger2021covid} further demonstrate the role of cultural, educational, and experiential factors in shaping vaccine attitudes, advocating for targeted and context-aware communication campaigns. These findings underscore the importance of integrating dynamic and context-sensitive approaches into vaccine hesitancy research and intervention planning.

While these methods have significantly expanded our understanding of vaccine hesitancy, it is also crucial to incorporate mathematical modelling to explore how people's attitudes toward vaccination have evolved over time and to explain the observed dynamics mechanistically. This study specifically aimed to collect daily information from a representative sample of Facebook's Active User Base, who were invited to participate in the survey to provide insights on various COVID-19-related factors, including symptoms,  test results, vaccination acceptance, isolation measures, and demographic information.  We model the temporal variations in vaccination attitudes by categorising individuals into three groups: those who are vaccinated, those who are still not vaccinated but accept the vaccine, and those who refuse it. We implement our models on data from the COVID-19 Trends and Impacts Survey (CTIS)~\cite{Salomon2021}, a cross-sectional survey developed by the Delphi group at Carnegie Mellon University in collaboration with Facebook. Our models are designed to investigate how interactions among individuals--categorised as vaccinated, accepted and refused--can influence attitudes toward vaccination over time.  We also examine the relationship between political affiliation at the state level in the United States and vaccination rates, exploring how political dynamics may shape and shift vaccination attitudes. Finally, we analyse how demographic factors, such as age and education level, impact vaccination behaviour, opinion changes, and the temporal dynamics of attitude shifts. These insights  comprehensively understand the socio-political and demographic factors driving vaccination trends and the evolution of public opinion.

The organisation of the paper is as follows: In Section~\ref{sec:materials_methods}, we present mathematical models that describe the transition between different vaccination classes (Vaccinated, Accepted, and Refused) with and without demographic structure. First, we present a model that describes the dynamics of these classes in an unstructured homogenous population; we refer to this model as the unstructured-Vaccinated-Accepted-Refused (uVAR) model. We then extend this model by including demographic structure; we refer to this model as structured-Vaccinated-Accepted-Refused (sVAR) model and it includes uVAR as a special case. We then outline relevant statistical methods, such as the Mann-Whitney U test and Principle Component Analysis (PCA), that we implement to assess our model's results. In Section~\ref{sec:Results}, the models are parametrised by fitting to survey data for each state in the USA. The influence of political affiliation on vaccination behaviour at the state level using the 2020 presidential election results and the influence across age and education groups is investigated. In Section~\ref{sec:discussion}, the paper concludes and quantified with a discussion of our findings and their implications.


\section{Material and Methods}\label{sec:materials_methods}
We want to understand how people change their attitudes towards vaccination over time. To this end, we develop a deterministic ordinary differential equation (ODE)-based  compartmental model to characterise the changing proportion of each class over time.  
The model parameters endow the rates of transitions between different population classes based on vaccination status. 
We estimate these parameters by calibrating them on USA survey data and finding their optimal values, which best fit the model to the data. 
We validate the model by comparing the time series of the proportion of the vaccinated population with survey and official data~\cite{Alvarez2021}. 
Moreover, to see how well the model describes the survey data and how well the survey data represents the actual trends, we calculated Pearson's correlations and the mean square error between survey data and model prediction. 
This is complemented by examining the statistical significance and performing the Mann—Whitney U test and principle component analysis. 

\subsection{The CTIS survey dataset}\label{sec:Dataset}
  The CTIS~\cite{Salomon2021} is an extensive national survey conducted daily from April 2020 through June 2022 to monitor the evolving COVID-19 pandemic in the USA. Our analysis focuses on the part of the survey that asks questions about vaccination status. 
The survey was conducted in a collaboration of the Delphi Research Group at Carnegie Mellon University and Facebook, along with a consortium of other academic institutions and public health partners~\cite{Salomon2021}. 
  Each day, around 40,000 people across the USA participated in the survey. 
  The sampling frame consisted of active Facebook users aged 18 years or older who had logged into their accounts within the past month. 
  These participants were recruited through advertisements randomly featured at the top of their Facebook news feeds. If users clicked on the advertisement, they were redirected to an external online survey administered by Carnegie Mellon University via Qualtrics, entirely outside the Facebook platform.
  The survey was available in multiple languages, including English, Spanish, Portuguese, Vietnamese, French, and Chinese, to reach a diverse nationwide sample.

 Carnegie Mellon University provided Facebook with anonymous respondent IDs to make the data more representative of the USA population. 
 Facebook then calculated survey weights for each participant based on demographic information available in their user profiles, such as age, gender, and location \cite{Barkay2020}. 
 The researchers at Carnegie Mellon University applied these weights when analysing the data to align the sample with USA population benchmarks. 
 These IDs maintained participant confidentiality, as Carnegie Mellon could not identify individual respondents, and Facebook never saw the survey response data. The weighting ensured the survey results were representative of the broader USA population. 
 The survey collected self-reported information related to COVID-19 symptoms, testing behaviours, vaccination status, attitudes toward vaccination, mask-wearing, underlying health conditions, mental health, employment, mobility patterns, and essential demographic characteristics. 
 Over 21 million USA Facebook users responded to the survey during the 810 days of data collection, providing invaluable insights into the ongoing public health crisis and how it evolved. 
 The large sample size and continuous daily data collection enabled the researchers to closely track trends and shifts in attitudes and behaviours throughout the pandemic. 
 The survey data was obtained by agreement with the CMU and Facebook; for more details on the survey, we refer the reader to~\cite{Salomon2021}. The biases related to demographic profiles, such as gender, age, and education, have been accounted for in the data through weighting adjustments. However, some sampling bias may persist regarding the likelihood of individuals choosing to participate in the survey. Typically, those who are more health-conscious and aware of vaccination tend to engage in such surveys. Nevertheless, a possible bias related to the willingness to complete the survey does not affect the study’s findings or conclusions.

\subsection{Data Preparation}\label{sec:Data_Preparation}
We classified the population into three classes based on their vaccination status to examine the dynamics of vaccination. 
Individuals who have received one or more doses of vaccine are considered ``Vaccinated (V)''.
Those individuals who have an appointment for vaccination or respond ``Yes, definitely'' and ``Yes, probably'' to the offer of a vaccine are considered ``Accepted (A)''.
The group of individuals who respond ``No, probably not'' and ``No, definitely not'' are labelled as ``Refused (R)''. We considered the data from $1^{st}$ January 2021 when vaccination started until the $20^{th}$ July 2022 which corresponds to 566 days.

The first step for utilising the survey data is cleaning and preparing the responses, using the method suggested by Álvarez \textit{et al.}~\cite{Alvarez2021}. The survey contains some questions requiring quantitative values, which sometimes have inconsistent values. 
Therefore, according to their methodology we removed responses with unusual values (greater than 100) for the quantitative questions about the number of days with symptoms, number of symptomatic contacts, number of people staying together, and years of education completed. 
Responses claiming all symptoms were also excluded. Some respondents may have exited the survey prematurely, failing to answer critical questions. Since we are interested in the frequency of responses to specific vaccination-related questions, we have only considered responses that answered those targeted questions. 
 The dataset contains responses grouped by state, however, some states have smaller sample sizes. Additionally, the number of individuals participating in the survey varies daily, as the number of invited participants who complete the survey correctly is not fixed. To ensure more robust daily data, we analyse the frequency of the three classes (V, A, R) instead of absolute counts for all states. Furthermore, each data point for a given day is computed using a 7-day moving average to smooth fluctuations and enhance reliability. This approach sacrifices the first week of data to smooth variability. We also utilized the weights provided by Facebook for each survey respondent, as previously described, to ensure that the results are representative of the broader USA population.

The resulting data is a time series for the classes of vaccinated, accepted, and refused individuals which we denote by $v(t)$, $a(t)$, and $r(t)$ respectively and where on any particular day $t$, $v(t)+a(t)+r(t)=1$. 

\subsection{Vaccinated-Accepted-Refused (VAR) model}
When a new vaccine is introduced into a population, people do not always respond in a binary way by accepting or refusing the vaccine, but can instead go through decision-making stages where they are influenced by other factors. We refer to this decision-making stage as vaccine hesitancy. Social interaction is one of the main drivers of the spread of infectious diseases and also of the adoption of interventions against infection such as vaccination. The demographic features of the individuals, such as sex, age, education, ethnicity, etc., powerfully shape this interaction. People who initially accept or refuse vaccines may change their decision when exposed to new information. Individuals who have accepted vaccination will either get vaccinated at a rate that depends on vaccine availability or may change their decision based on their interaction with individuals who have refused vaccination. Moreover, as time progresses, the positive feedback from vaccinated individuals and the vaccine availability will boost the vaccination rate when many individuals have accepted vaccines.
On the other hand, an individual who initially refused a vaccine may change their decision when interacting with vaccinated individuals or those who have accepted vaccines.
In a well-mixed homogeneous population, the interaction rate can be quantified using the mass-action law~\cite{erdi1989}, where the transition rates from one class to another are proportional to the frequency of interacting classes.


First we present the model that describes the dynamics of these classes in an unstructured homogenous population. We refer to this model as the unstructured-Vaccinated-Accepted-Refused (uVAR) model. Then, we extend the model by including the demographic structure. We refer to this model as the structured-Vaccinated-Accepted-Refused (sVAR) model.

\begin{table}[htp]
\caption{Summary: Notations for the uVAR model, and their definitions and descriptions.} 
\centering  
\begin{tabular}{p{2cm} p{2cm} p{10cm}} %
\hline  
 Notation & Definition & Description 
\\ 
\hline
$V$
& $\in [0,1]$
& The proportion of vaccinated individuals: Fraction of the population who has received at least one dose of any vaccine or has booked an appointment.
\\
$A$ 
& $\in [0,1]$
& The proportion of accepted: Fraction of the population who are not vaccinated responded ``Yes,
definitely'' and ``Yes, probably'' to the offer of a vaccine. 
\\
$R$ 
& $\in [0,1]$  
& The proportion of refused individuals: The fraction of the population who are not vaccinated responded  ``No,
definitely'' and ``No, probably'' to the offer of a vaccine. 
\\
$\alpha$ 
& $\in [0,1]$  
& Baseline vaccine rate at the beginning of vaccine rollout.
\\
$\gamma$ 
& $\in [0,1]$  
& Parameter that represents the increase in the availability of vaccines with time.
\\
$\beta$ 
& $\in [-1,1]$  
& Net rate of influence through which refused individuals change their opinion to accept vaccines based on their interaction with individuals in the vaccinated or accepted classes.
\\
$A_r$ 
& $\in [0,1]$  
& Fraction of individuals who have accepted the vaccine but remained unvaccinated throughout the campaign. Referred to as `residuals accepted' as indicated by the subscript $r$.
\\
$R_r$ 
& $\in [0,1]$  
& Fraction of individuals who have refused the vaccine and have never changed their decision throughout the campaign. Referred to as `residuals refused' as indicated by the subscript $r$.  
\\
\hline  
\end{tabular} 
\label{tab:VAR-params}
\end{table}
 

\subsubsection{uVAR Model}
Consider a well-mixed homogeneous population in which individuals are grouped according to their vaccination status: Vaccinated ($V$), Accepted ($A$), and Refused ($R$).
 By the end of a vaccination campaign, some individuals refuse vaccines due to unchanged concerns like safety, vaccine type, or government distrust. In contrast, others accept but remain unvaccinated for some reasons, for example, health issues, availability, or logistical barriers. This can be seen in the data under consideration and also official data reported by the USA Centers for Disease Control and Prevention (CDC) COVID Data Tracker~\cite{CDC_COVID_Tracker}.   
Therefore we assume that only the proportion $A-A_r$ of individuals in accepted and the proportion $R-R_r$ in the refused classes would have the potential to move between classes, with $A_r$ and $R_r$ denoting residual accepted and refused individuals, respectively. The structure of the uVAR model is shown in Figure~\ref{fig:uVAR-diagram} and the differential equations describing the flow between each compartment are given by:

\begin{figure}
    \centering
    \includegraphics[width=1\linewidth]{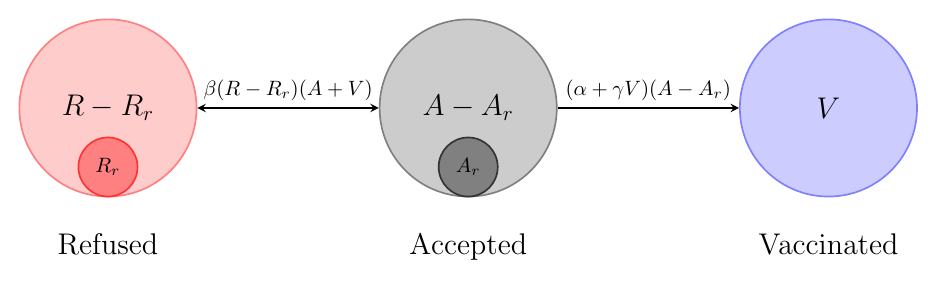}
    \caption{Flow diagram of states in Vaccinated-Accepted-Refused model and transition rates between them.}
    \label{fig:uVAR-diagram}
\end{figure}

\begin{equation}\label{equ: uVAR-model}
  \!\begin{aligned}  
       \frac{dV}{dt}&=(\alpha +\gamma V)(A-A_r) \\  
       \frac{dA}{dt}&=-(\alpha +\gamma V)(A-A_r)+\beta (R-R_r)(A+V)\\
    \frac{dR}{dt}&=-\beta (R-R_r)(A+V).  
    \end{aligned}
\end{equation}  
Here, the vaccination rate depends on the number of accepted individuals, which will actively seek vaccination ($A-A_r$). It also depends on the vaccination availability or vaccine rate, which we have modelled as $\alpha+\gamma V$. As the vaccine rollout progressed, the availability of the vaccine increased during the early stages. Here, $\alpha$ represents the vaccination rate at the beginning of vaccination campaign when vaccines was very restricted. The term $\gamma V$ is a proxy to allow this initial rate to grow in time in proportion to the number already vaccinated. Eventually this rate becomes large, which is consistent with full availability of the vaccine for any individual who wishes to become vaccinated.

We assume that the refused individuals who are susceptible to change, ($R-R_r$), transition to the accepted class by being influenced by those in the vaccinated and accepted classes. This influence can take many forms, including direct communication, and the contribution of vaccinated individuals to the statistics on the efficacy and safety of those vaccinations which are subsequently reported in the media. It is also possible for accepted individuals to become refused by interaction with refused individuals yielding a term proportional to $AR$ in the differential equation for $R$. However, we assume this effect to be absorbed into the value of $\beta$ which is always found to be positive and represents the net rate of influence from refused to accepted. While more complex models can be designed, our model represents the simplest formulation that fully accounts for the observed dynamics in these three groups in the survey data. 

We assume that the rates $\alpha$ and $\gamma$ are in $[0,1]$, while $\beta$ can range from $[-1,1]$, with its sign indicating the direction of flow. A positive $\beta$ represents the flow from $R$  to $A$, while a negative $\beta$ indicates flow from $A$ to $R$. This flexibility allows for the modelling of opinion changes in both directions, as individuals may shift their stance to accept the vaccine after initially refusing it, or conversely, change from acceptance to refusal due to various factors such as negative feedback, concerns, or vaccine availability. Since the parameters $A_r$ and $R_r$ are proportions of the population, they are also assumed to be between $0$ and $1$.

\subsubsection{sVAR Model}
We are interested in the impact of age and level of education on the influence between groups. For this, we now extend the nVAR model to include demographic structures. If we have $n$ demographic groups, then we obtain a model with $3n$ compartments, accounting for the three vaccination states.
Given that all compartments interact with  each other, then the dynamics can be described by the following model:

\begin{equation}\label{equ: sVAR-model}
    \!\begin{aligned}  
       \frac{dV_i}{dt}&=(\alpha_i +\gamma_{i} V_i)(A_i-A_{r,i}) \\  
       \frac{dA_i}{dt}&=-(\alpha_i +\gamma_{i} V_i)(A_i-A_{r,i})+ (R_i-R_{r,i})\sum_{j=1}^n\beta_{ji}(A_j+V_j)\\
    \frac{dR_i}{dt}&=-(R_i-R_{r,i})\sum_{j=1}^n\beta_{ji}(A_j+V_j),    
    \end{aligned}
\end{equation}
where $V_i$, $A_i$, $R_i$ are the fractions of vaccinated, accepted, and refused in group $i$, respectively, and $A_{r,i}$, $R_{r,i}$ are the residual fractions of accepted and removed in group $i$, respectively, for $i\in\{1,2,\dots,n\}$. Here, $\beta_{ji}$ represents the rate of influence of demographic group $j$ on demographic group $i$, and the other parameters have the same meaning as in the uVAR model, except that $i$ denotes the associated group. Note that $\sum_{i=1}^n (V_i+A_i+R_i)=1$ and that for $n=1$, this model reduces to the uVAR model as a special case. 

\subsection{Statistical Methods}
We find the best fit to our models by minimising the mean square error (MSE) given by
\begin{equation}
    \text{MSE}=\frac{1}{m}\sum_i^n(y_i-\hat{y}_i)^2.
\end{equation}
where $Y=(y_1,y_2,\dots,y_m)$ is the observed survey data (i.e., the time series data) and $\hat{Y}=(\hat{y}_1,\hat{y}_2,\dots,\hat{y}_m)$ is the prediction from our model. For demographic group size $n$, we have $m=3nT$ where $T=566$ is the number of days of data. Minimisation of MME is achieved by the differential evolution (DE) optimisation method (see Section \ref{sec: optimisation} for details). Fits were performed for each of the 50 states in the USA independently, yielding 50 values for the model parameters.

To investigate the influence of political affiliation, we categorise states as Republican and Democratic based on the 2020 presidential election results, and conducted a Mann-Whitney U test and analyzed the corresponding $p$-value (see Section \ref{sec:MWU-test} for details). The Mann-Whitney U test assesses whether two groups differ in their central tendencies. The $p$-value derived from the test indicates the statistical significance of the difference. For example, a $p$-value less than $0.05$ typically suggests that the two groups
differ significantly. To visualise and compare how these parameters vary across groups, we created boxplots for each model parameter, categorised by different groupings such as political affiliations or demographic characteristics. 
 
The uVAR model involves five parameters, while sVAR model involves $n^2+4n$ parameters, where $n$ is the number of demographic groups. 
We performed Principal Component Analysis (PCA) on the space of parameters.
PCA is a dimensionality reduction method that identifies directions of maximum variance to simplify data representation. Reducing higher dimensioned parameter spaces to a two dimensioned parameter spaces, we classified states, effectively labelled by their political affiliations. We used the first two principal components derived as eigenvectors of the covariance matrix corresponding to its two largest eigenvalues. These components capture the directions of maximum variance, enabling simplified data visualisation. If $X_{\text{std}}$ is the standardised matrix associated with the input data matrix $X$ whose columns are made-up of the parameters vectors, then the data transformed into two-dimensional space is given by 
\begin{equation}
 Z=X_{\text{std}}\begin{bmatrix}
     PC_1\\
     PC_2
 \end{bmatrix}   
\end{equation}
where $PC_1$ and $PC_2$ are the principle eigenvectors of the covariance matrix $\Sigma$ of the matrix $X_{\text{std}}$. For details, we refer to Section~\ref{sec:PCA} and references cited there.

We employed the bootstrapping method to enhance the reliability and robustness of our statistical analyses. Bootstrapping is a resampling technique that generates multiple datasets by randomly sampling with replacement from the original data, allowing us to create a larger representative dataset~\cite{debeire2024bootstrap}. This approach enabled us to perform more comprehensive statistical tests, calculate $p$-values more confidently, and apply the PCA algorithm effectively. To implement bootstrapping, we repeatedly selected random subsets of states (in our analysis, we select 5 random states to form a subset), ensuring that the subsets maintained the same labels as their original groups (e.g., Democratic states). These selected states were aggregated to form new resampled datasets. By performing this process numerous times (5000 times in our analysis), we generated a larger synthesised dataset that preserved the original group labels, enabling us to conduct more robust statistical analyses.

To assess how well our survey data and mathematical model capture trends observed in official government-reported data, we used the Pearson correlation coefficient, which is defined by 
\begin{equation}
    r=\frac{\sum_i^n (x_i-\Bar{x})(y_i-\Bar{y})}{\sqrt{\sum_i^n (x_i-\Bar{x}})^2\sqrt{\sum_i^n (y_i-\Bar{y}})^2},
\end{equation}
for samples $x=(x_1,x_2,\dots,x_n)$ and $y=(y_1,y_2,\dots,y_n)$ of two random variables and where $\Bar{x}$ and $\Bar{y}$ represent the sample means of $x$ and $y$, respectively.  The values of $r$ range from $-1$ to $1$, where $r=1$ ($r=-1$) indicates a strong positive (negative) correlation between $x$ and $y$, and $r=0$ indicates no linear relationship between the two variables.

\section{Results}\label{sec:Results}
\begin{figure}[th]
    \centering
    \includegraphics[scale=.3]{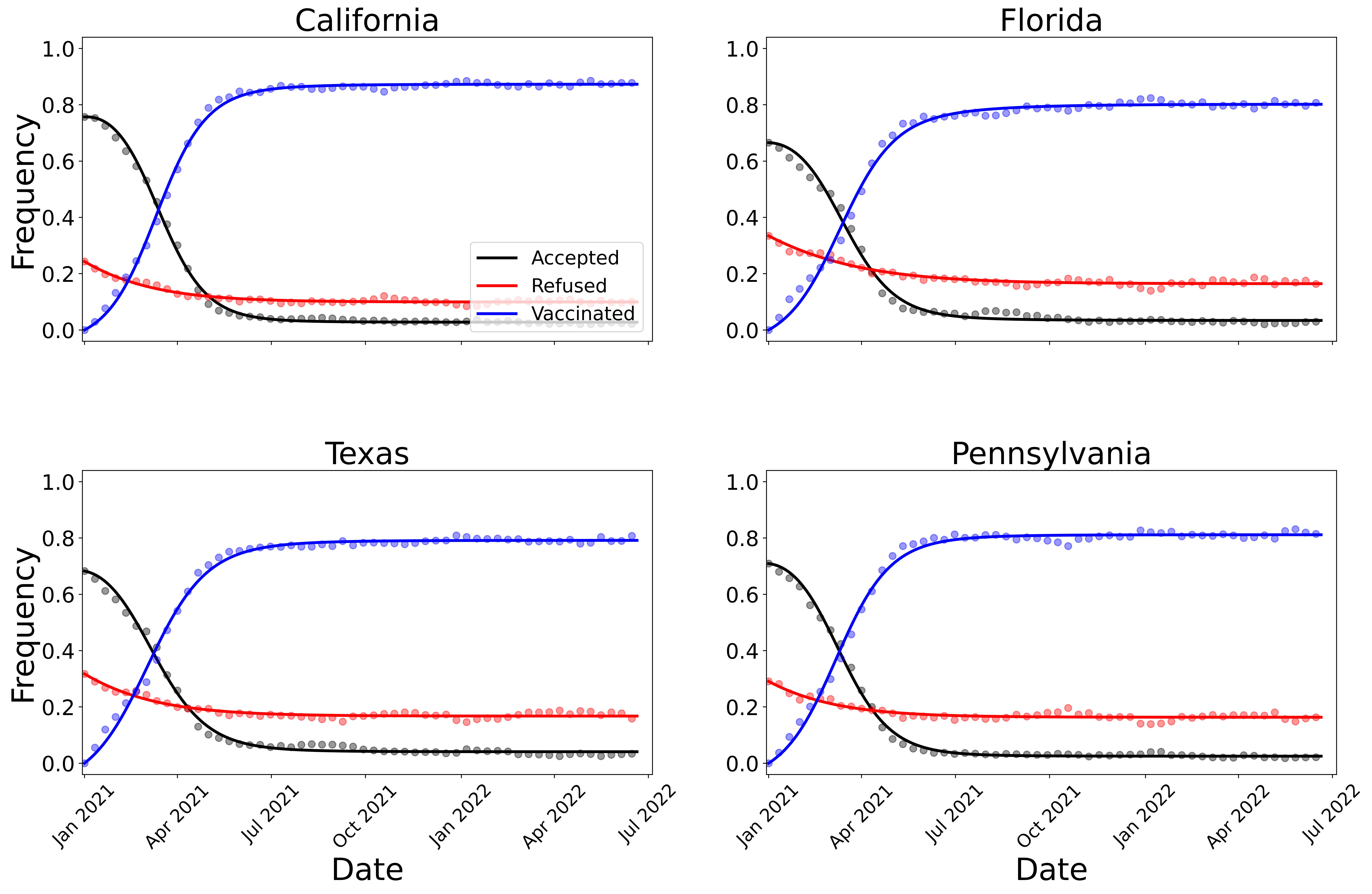}
    \caption{Comparison of the uVAR model predicted curves to data from the survey for four states (California, Florida, Texas, and Pennsylvania) (The dashed lines are the solution curves of uVAR Model~\eqref{equ: uVAR-model}, and the dots show the survey data. Blue, black, and red indicate vaccination, acceptance, and refused individuals, respectively. The complete results for all 50 states are presented in Figure \ref{fig:fitVAR_all_states}).
    }
    
    \label{fig:fitVAR4states}
\end{figure}

\begin{figure}[th]
    \centering
    \includegraphics[scale=.5]{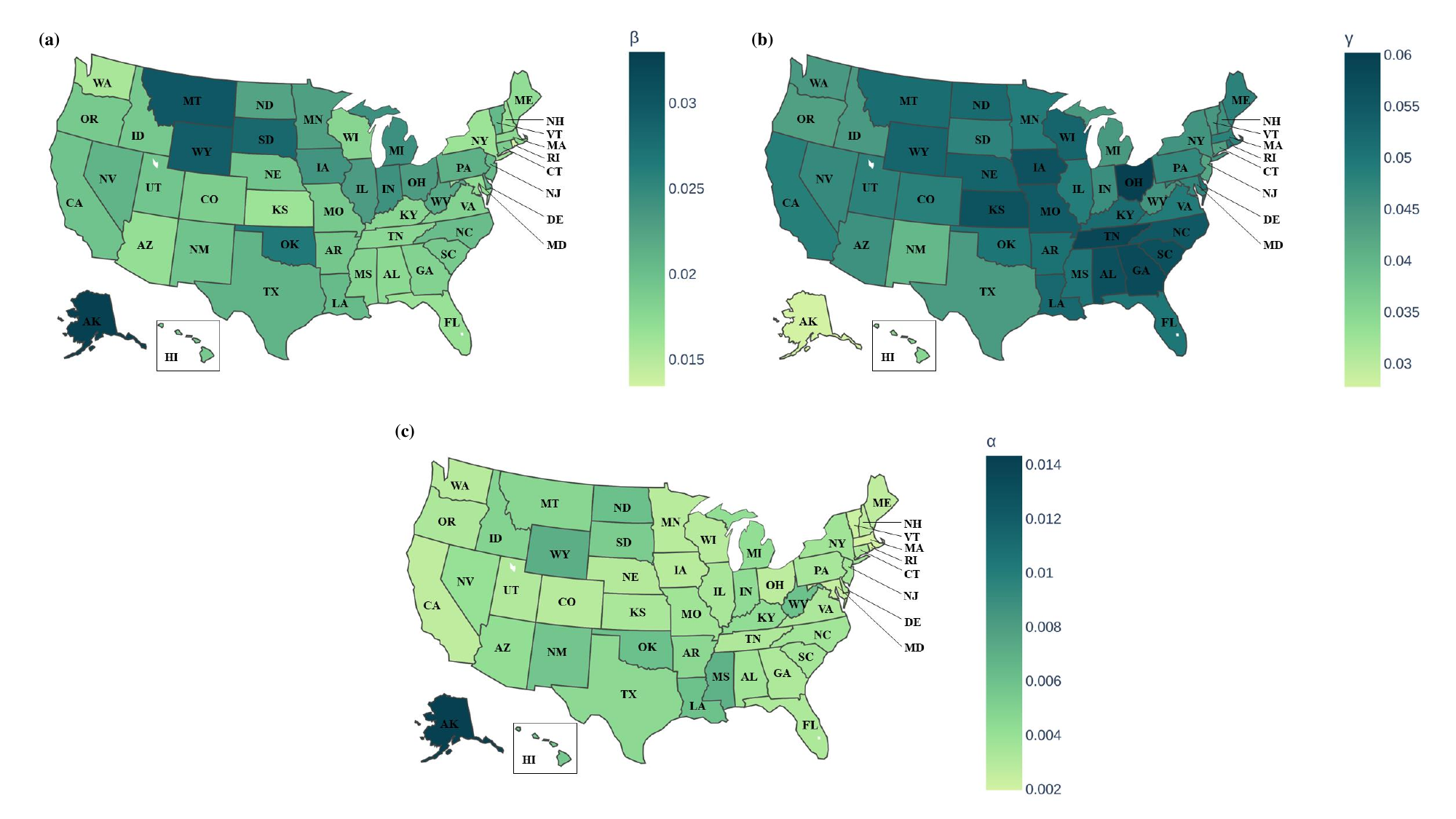}
    \caption{The optimised parameter values for each USA state (The states are represented by their state codes provided in Table~\ref{tab:parameters}). The density of the colour on the map shows the optimised values of parameters $\beta$, $\gamma$ and $\alpha$ in (a), (b), and (c), respectively. The values are achieved by fitting uVAR Model~\eqref{equ: uVAR-model} to state-level data. The exact values of the parameters can be seen in Table \ref{tab:parameters}).}
    \label{fig:VAR_params_all_states}
\end{figure}

In this section, we apply our models to investigate vaccination hesitancy at the state-level and national-level using the data discussed in Section~\ref{sec:Data_Preparation}.  The uVAR Model~\eqref{equ: uVAR-model} demonstrates strong fits to the data, as shown in Figure \ref{fig:fitVAR4states} for four states. The complete results for all $50$ states are displayed in Figure \ref{fig:fitVAR_all_states} in the Supplementary Material, which demonstrates the uVAR model’s excellent performance in describing the data. Figure \ref{fig:VAR_params_all_states} shows the optimized parameters $\beta$, $\alpha$, and $\gamma$ for the models featured in all states (see Table~\ref{tab:parameters} for values of the parameters). 
These results highlight how our model can describe the survey data and people’s behaviour toward COVID-19 vaccination. 
In January 2021, the frequency of vaccinated people was low across all states, with almost no one vaccinated. Initially, the proportion of accepted group is large and declines as vaccines become more available and the number moving into the accepted class from refused declines.  The proportion of refused class also decreased, indicating that some people who were previously hesitant later decided to take the vaccine. 

Across all states, the model parameter $\beta$ is positive, meaning more people changed their stance from refused to accepted than vice versa, leading to a net flow from refusal to acceptance. When the vaccines became available, there was a sharp uptick in vaccination frequency as people were eager to get their shots immediately. 
At the end of the pandemic, the population was largely divided into two groups: those vaccinated and those who refused to get vaccinated. 

\subsection{Comparison between Official, Survey, and Model Predictions}
The models are optimised using survey data. 
We compared the survey and fitted model results to official USA vaccination data from the USA Center for Disease Control and Prevention (CDC) COVID Data Tracker~\cite{CDC_COVID_Tracker}. CDC reports the percentage of the population that have received at least one vaccine dose from December 14, 2020, to May 9, 2023. We only considered the time frame from January 1, 2021, to July 20, 2022, for an equivalent comparison period across the survey, model, and CDC data. 

\begin{figure}
    \centering
    \includegraphics[scale=.20]{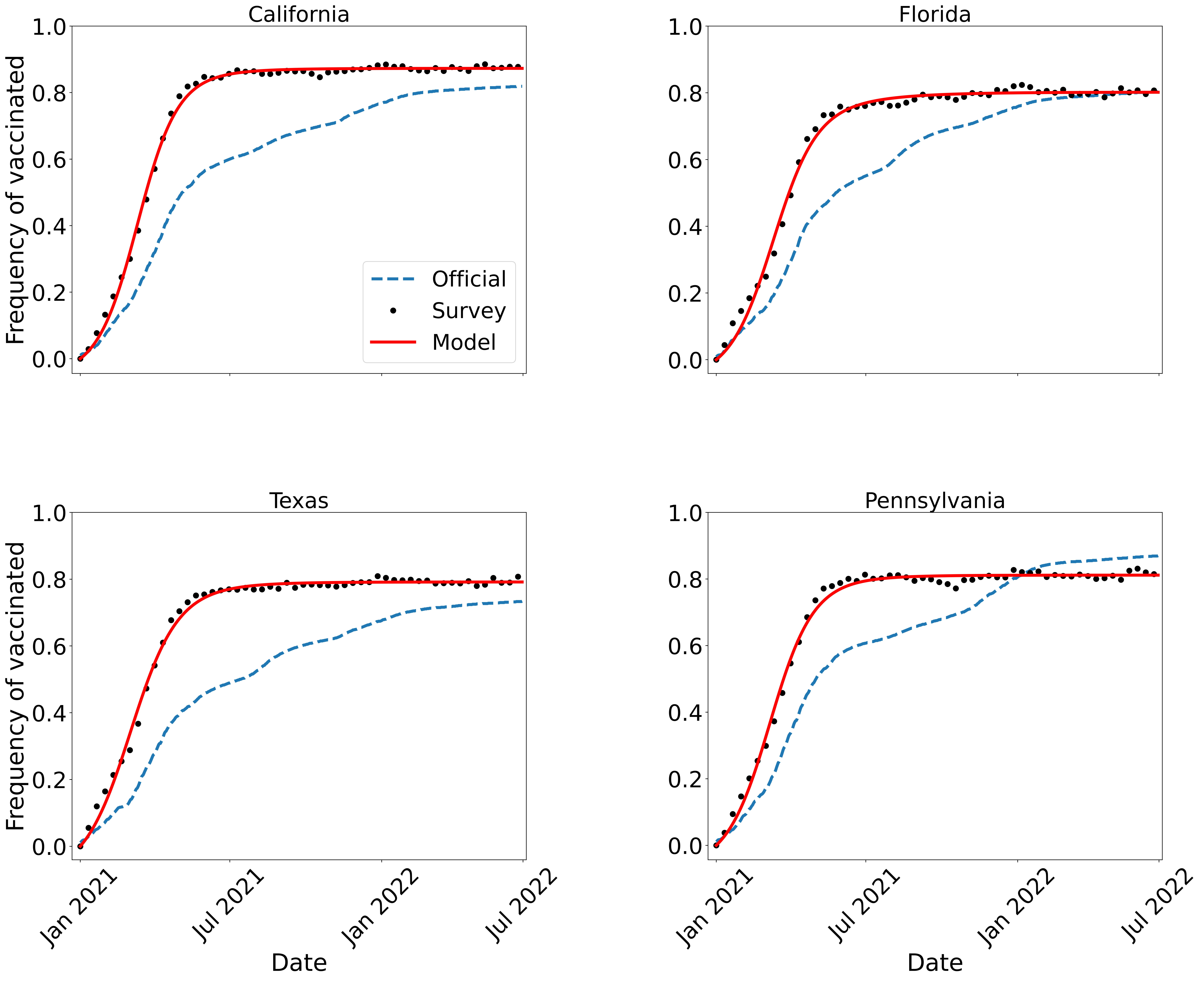}
    \caption{Comparison of the time-evolution of the proportion of vaccinated individuals predicated by uVAR Model~\eqref{equ: uVAR-model} with survey and official data of four states in the USA (California, Florida, Texas, and Pennsylvania).}
    \label{fig:VAR-Survey-Official}
\end{figure}

Figure~\ref{fig:VAR-Survey-Official} shows how vaccination frequency changed over time in four USA states: California, Florida, Texas, and Pennsylvania. It compares the official proportion of vaccinated residents to survey and model prediction from January 1, 2021, to  July 20, 2022. In all four states, the model closely matches the survey data, as expected, since the model is optimised based on these surveys. Generally, the three data sources follow roughly similar upward trends in each state. However, the survey data consistently overestimates official vaccination rates and this is particularly true earlier in the time series. By July 20, 2022, which is the end of the period for which we have data, the gap closes significantly, with all estimates ending at around 80 percent fully vaccinated. 

One key factor contributing to this discrepancy is the delay in reporting and updating official vaccination data. As noted by the CDC \cite{CDC_COVIDVaxView}, official records lag behind updates provided on the Respiratory Illnesses Data Channel by several days. This delay arises from the time required to review data across different geographic and sociodemographic groups and to synthesize the findings. Consequently, temporary mismatches may occur, where survey data indicate higher vaccination rates than those captured in official records.  However, given that our data represent a seven-day aggregation, reporting delays alone do not fully account for the observed discrepancy, especially during the middle of the study period. Another plausible explanation is a selection bias inherent in survey participation. Individuals who engage in such surveys are more likely to be those who are particularly concerned about COVID-19 and, therefore, more inclined to receive the vaccine. In contrast, those who perceive COVID-19 as a lesser threat may be less motivated to participate in surveys or related social initiatives. This self-selection bias can lead to an overrepresentation of vaccinated individuals in survey data, particularly during the early stages of vaccine distribution, further amplifying the divergence between survey-based estimates and official records.

\begin{figure}
    \centering
    \includegraphics[scale=.30]{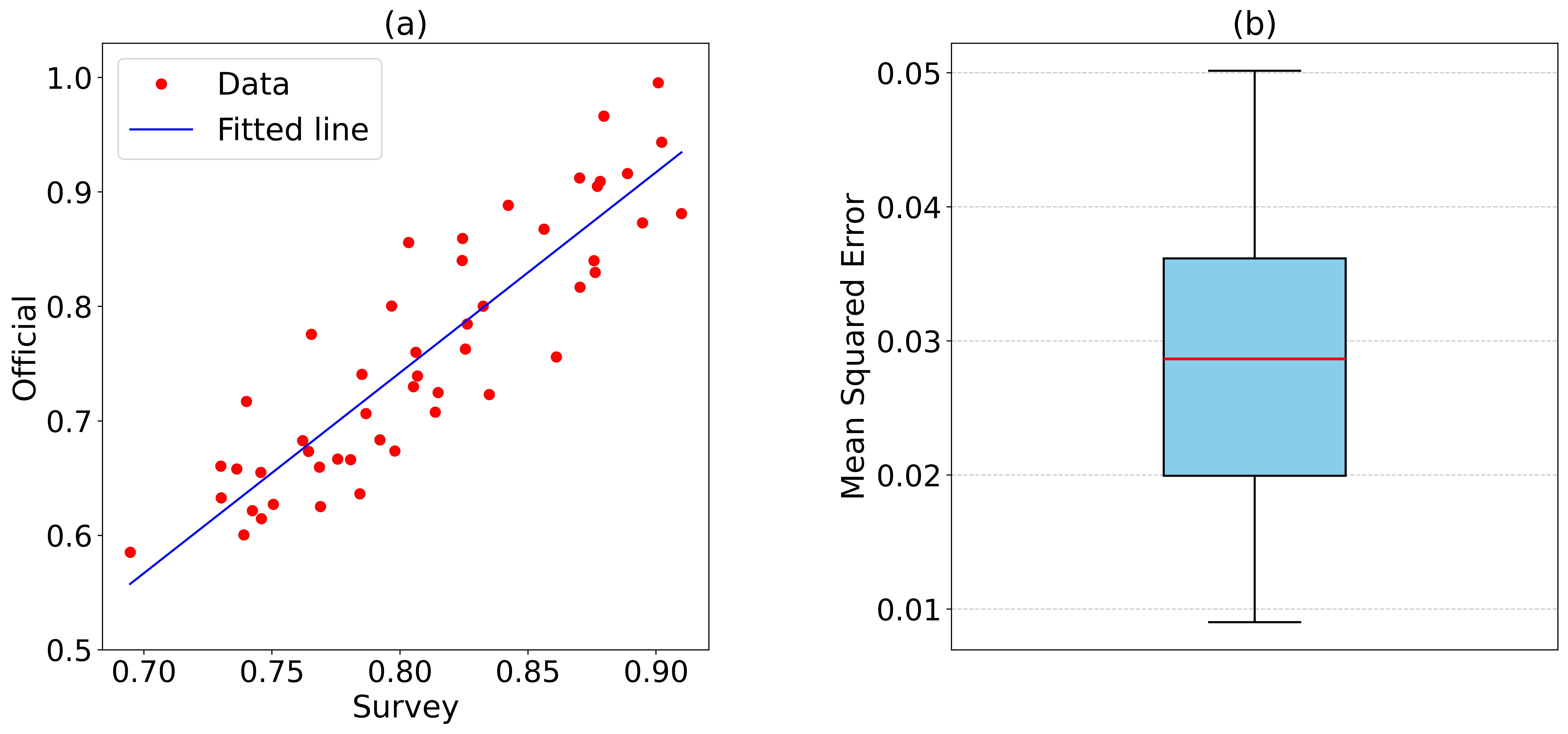}
    \caption{Comparison between the proportion of vaccinated on  survey data and official statistics (a) The dots are the final fraction of vaccinated data on July 20, 2022 plotted against one another, and the blue line is a linear fit of the form $y=ax+b$ where $a=1.715$, $b=-0.659$ and there is a Pearson Correlation($r=0.89$) and MSE=$0.007$. (b) The box plot showing MSE between official and survey data for each state from  January 1, 2021, until July 20, 2022. The mean of MSEs is $0.028$.}
    \label{fig:MSE_Pearson_correlation}
\end{figure}

To further assess whether the survey and model adequately capture the official vaccination data, we closely examined the final vaccination rates reported for each state as of July 20, 2022 (see Figure \ref{fig:Vaccination_survery_model_official} for individual state values) and calculated the MSE)between them over the period from January 1, 2021, to July 20, 2022. Figure \ref{fig:MSE_Pearson_correlation} provides a direct comparison of the alignment between official data and survey data by plotting and computing both the MSE and Pearson correlation coefficient ($r$) for the final vaccination proportions and the MSE between time series for each state. Figure \ref{fig:MSE_Pearson_correlation}(a) shows a notably low MSE ($0.007$) and a high Pearson correlation ($0.89$) between the survey projections and official data, reinforcing our earlier claim that the final vaccination fraction estimated from the survey (and consequently the model optimized based on it) closely matches the official records. This suggests that the survey data offer a reasonable approximation of statewide vaccination coverage, making the model a useful mathematical tool for analysing broader vaccination trends and behaviours. However, Figure \ref{fig:MSE_Pearson_correlation}(b) indicates that the average MSE between the survey data and official records across all states is higher, at $0.028$. This discrepancy arises because, while the final fractions align well, the survey data tend to underestimate the official figures when considering trends over time.

\subsection{How Political Affiliation Can Affect Vaccination Behaviour}
The political landscape of the USA consists of states that tend to vote for either Republican or Democratic candidates in significant elections like the presidential race and congressional midterms. States are often categorised as “red states” that predominantly vote Republican or “blue states” that largely vote Democratic. This political affiliation at the state level based on voting patterns provides insight into the underlying ideological differences between states. For example, red states favour more conservative policies aligned with Republican values, while blue states lean towards more liberal views endorsed by Democrats. 
Therefore, examining people’s behaviour toward vaccination between Republican-affiliated red and Democratic-affiliated blue states can highlight how partisan identities impact attitudes and decision-making regarding critical preventative health measures. Analysing the interplay between a state’s partisan political affiliation and COVID-19 vaccination behaviour will thus provide a greater understanding of how ideology and identity politics influence life-and-death outcomes during a public health crisis. For this, we aggregated data from the states for each political affiliation. 

\begin{figure}
    \centering
    \includegraphics[scale=.25]{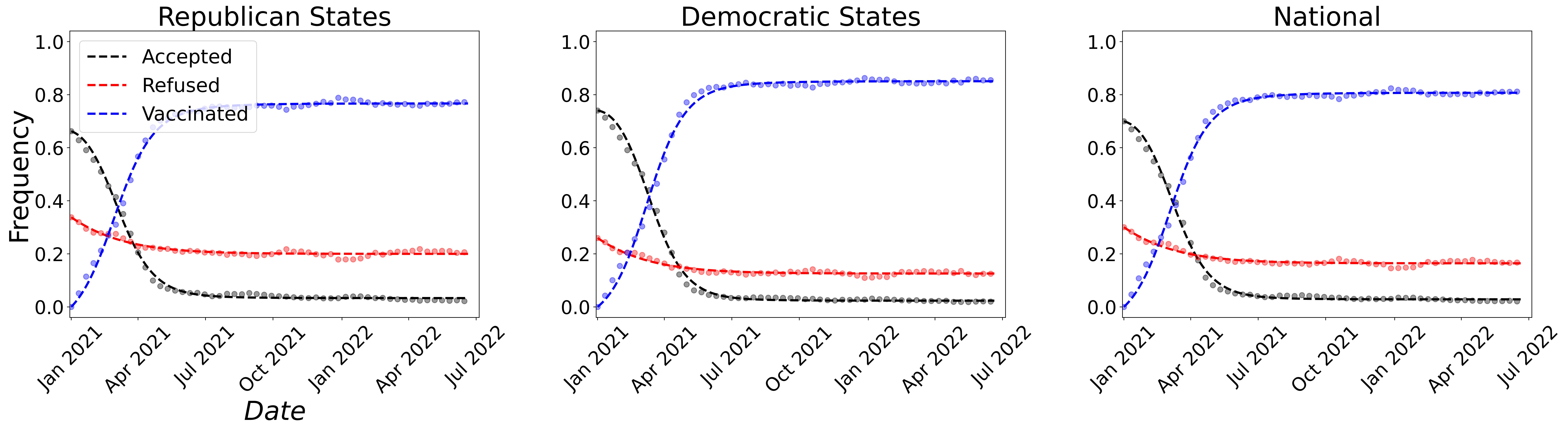}
    \caption{Comparison of the uVAR model's fits to data from the survey for Republican states, Democrat states, and the whole country (The dashed lines show the results from uVAR Model \eqref{equ: uVAR-model}, and the dots show the survey data. Blue, black, and red indicate vaccination, acceptance, and refused individual respectively).}
    \label{fig:VAR_political_national}
\end{figure}

Figure \ref{fig:VAR_political_national} illustrates the temporal progression of vaccination rates across states with the majority of votes for Republican and for Democratic parties in the 2020 presidential election and for the national average. The figure reveals that, on average, by the end of the period under consideration, states with Democratic leadership attained  a higher fraction of their population vaccinated ($0.83$) compared to Republican states ($0.75$). Figure \ref{fig:VAR_scatter_plots} displays the uVAR Model~\eqref{equ: uVAR-model} parameters for each state, with states color-coded by political affiliation based on the 2020 presidential election results (see Figure~\ref{fig:political_affiliation}). As shown, state affiliation correlates with vaccination behaviours. Specifically, the parameters $\alpha$, $A_r$, and $R_r$ are categorised by political orientation. On average, Republican states exhibit higher values for all parameters, reflecting faster opinion changes between different groups. 

\begin{figure}
    \centering
    \includegraphics[scale=.38]{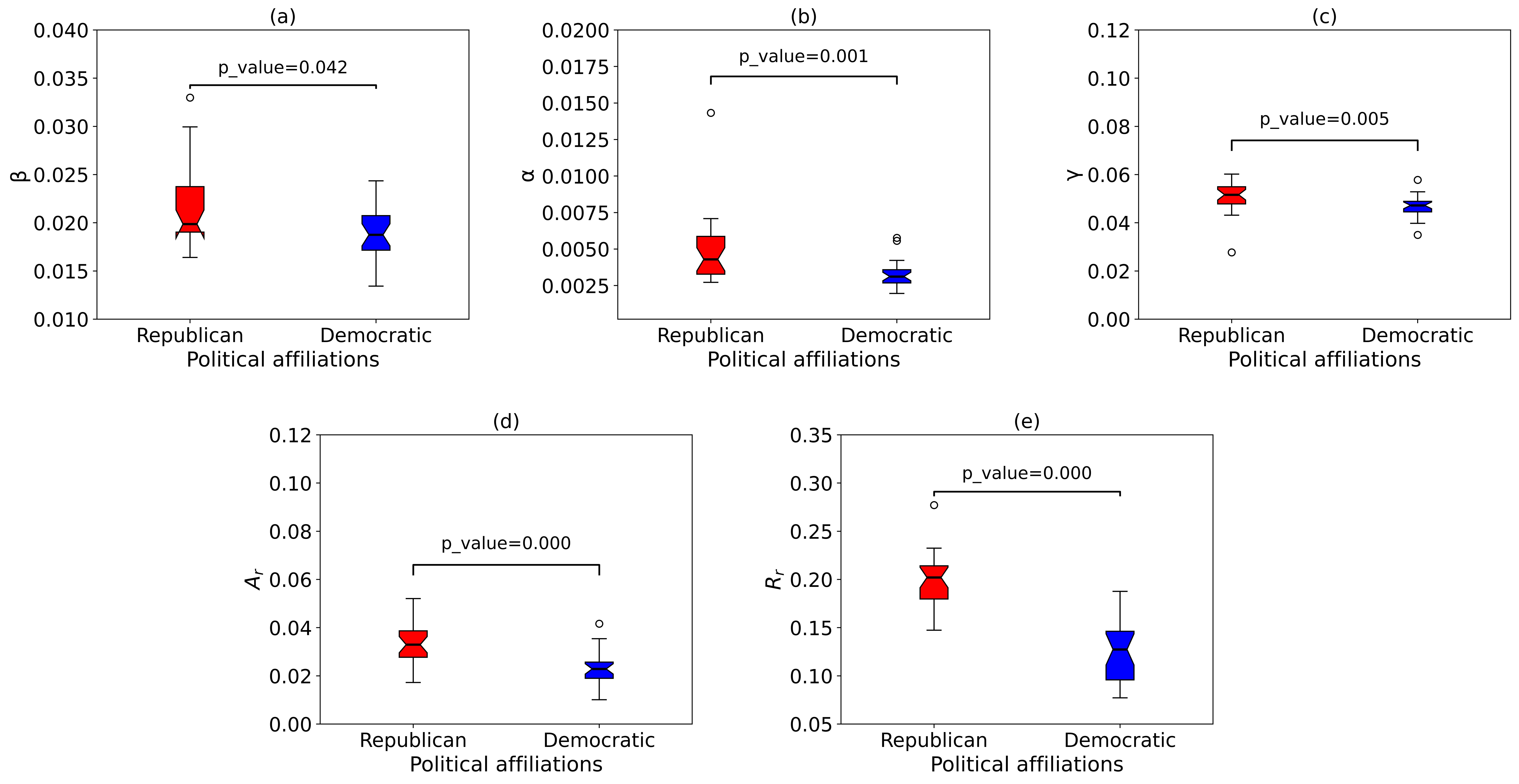}
    \caption{ Boxplot of parameters $\beta$, $\alpha$, $\gamma$, $A_r$ and $R_r$ for different states based on the uVAR Model presented in Equations~\eqref{equ: uVAR-model} (The red boxes are plotted using Republican states, while the blue corresponds to Democratic states. The two sets of states are also compared with the help of the Mann-Whitney U test; the $p$-value calculated from this test is displayed on each subplot).}
    \label{fig:VAR_boxplotpara}
\end{figure}

\begin{figure}
    \centering
    \includegraphics[scale=.35]{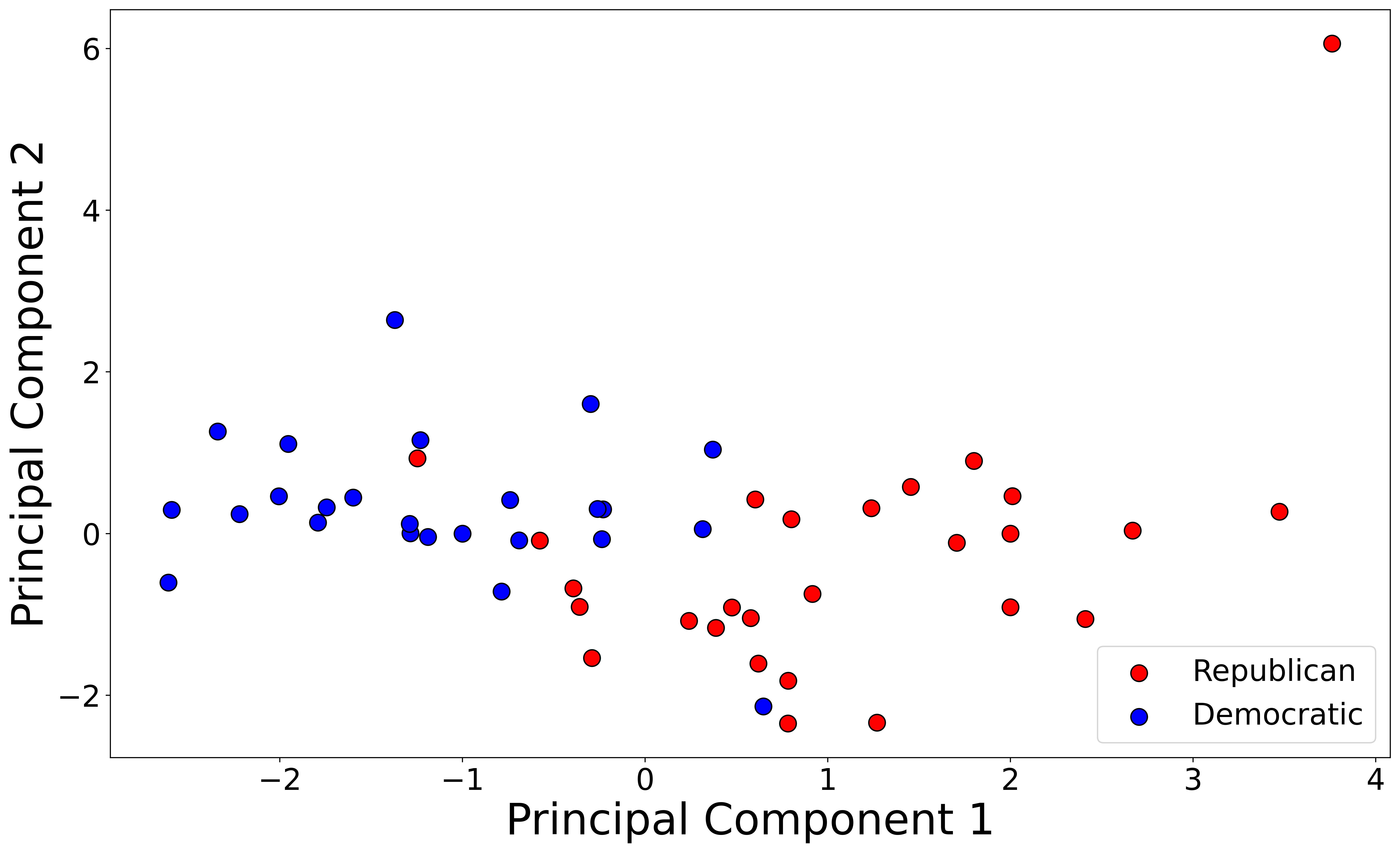}
    \caption{PCA clustering of states based on five vaccine model parameters, highlighting distinct political affiliations with Republican states in red and Democratic states in blue.}
    \label{fig:PCA_state_wise}
\end{figure}

\begin{figure}
    \centering
    \includegraphics[scale=.48]{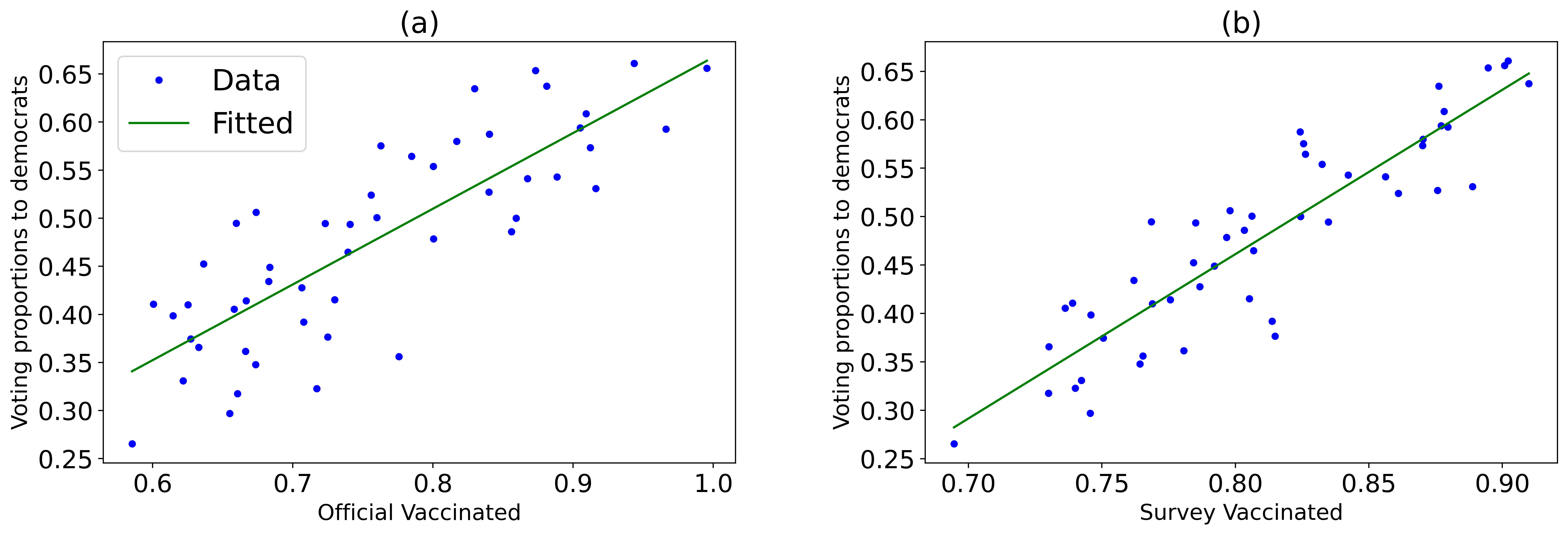}
    \caption{Comparison of the proportion of voted with the proportions of vaccinated individuals in the survey and official statistics (line $y=ax+b$, a solid green line, is fitted to the data in blue dots).  The Pearson correlation ($r$) between the proportions and vaccinated proportions is also computed.  An individual is counted as vaccinated if they have received at least one dose based on official data and survey. (a) $a=0.787$, $b=-0.120$ and $r=0.82$ (b) $a=1.696$, $b=-0.896$ and $r=0.90$).}
    \label{fig:PC_voter_vaccinated}
\end{figure}

The parameters for each state are summarised in the box plots in Figure~\ref{fig:VAR_boxplotpara}, indicating that the parameters for the Republican states have higher medians when compared to those using data from the Democratic states. On average, more variability can be seen in the parameters for Republican states. The lower values of parameters in Democratic states can be associated with the higher initial proportions of the accepted individuals. The white circles represent the outliers in the data and correspond to the state of Alaska. The non-zero parameters $A_r$ and $R_r$ indicate that a significant number of people never received a vaccine. To determine whether these observed patterns reflect real effects or are a result of random fluctuations we use the statistical Mann-Whitney U test. 

The $p$-values calculated from the Mann-Whitney U test for the set of parameters \{$\beta$, $\alpha$, $\gamma$, $A_r$, $R_r$\} between Republican and Democrat groups is \{0.042,0.001,0.005,0.000,0.000\}. Since all of the $p$-values are smaller than the 0.05 significance level, there is strong statistical evidence to reject the null hypothesis of no difference between Republican and Democratic states. Specifically, the very low $p$-value for $\alpha$, $A_r$, and $R_r$ suggests a highly significant difference between Republicans and Democrats, with the distributions unlikely to occur if there was no true difference. In Figure \ref{fig:PCA_state_wise}, we performed Principal Component Analysis (PCA) on all five model parameters to reduce their dimensionality, capturing the most significant variations in just two principal components. These two components together explain the percentage of the total variance, ensuring a more interpretable representation of the data. The figure shows that the five parameters of the vaccine model go a long way towards clustering all the 50 states based on their political affiliation. This clustering highlights the potential influence of political factors on the model parameters.

In Figure~\ref{fig:VAR_boxplotpara}, the smaller values of $A_r$ and $R_r$ of Democratic states compared to the Republican states suggest a higher proportion of vaccinated individuals in these than the Republican  states, which can also be seen in Figure~\ref{fig:VAR_political_national} where the proportion of vaccinated individuals in Democratic states is $0.83$ compared to that of Republican states that is $0.75$. However, a question may arise  regarding the method used to label states as Republican and Democratic; that is, based on the 2020 presidential election, which may not effectively represent the entire population. For example,  a slightly over 50\% ratio for one political party would place it in that category. To investigate this, we also used the actual proportion of voters for each state and performed Pearson correlation with the proportion of vaccinated individuals (Figure~\ref{fig:PC_voter_vaccinated}). We observed a strong positive correlation between the proportion of vaccinated individuals and Democratic voters, which supports our observation in survey data based on categorising states according to the 2020 presidential elections. Also For Republic states there is a negative correlation between the proportion of voted and the proportion of vaccinated people.

\subsection{Bootstrapping}
\begin{figure}
    \centering
    \includegraphics[scale=.3]{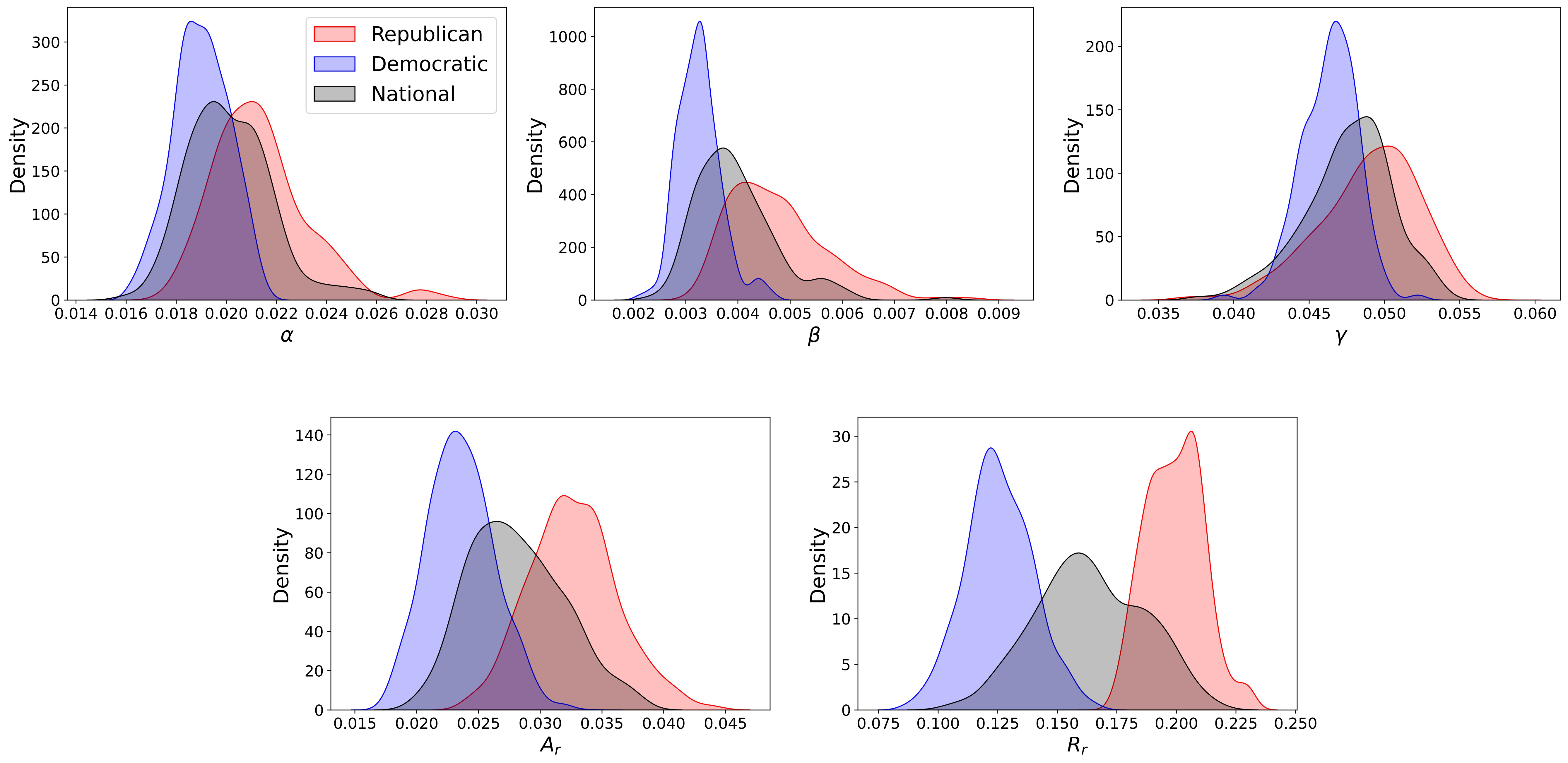}
    \caption{Bootstrapped distribution of (a) baseline vaccination rate, $\alpha$ (b) vaccine acceptance/refusal, $\beta$ (c) rate of influence of vaccinated on accepted class $\gamma$ (d) residual proportion of accepted, $A_r$ (e) residual proportion of refused, $R_r$, all estimated by implementing uVAR Model~\eqref{equ: uVAR-model} on the datasets of states separately based on political affiliations (Democratic or Republican) and all of them together (National) (In each group, we randomly selected $5$ states and aggregated them as a dataset, then we optimised our model and found parameters. We did it for each group $5000$ times).  
    }
    \label{fig:VAR-Bootstrapped}
\end{figure}

Fitting the uVAR Model~\eqref{equ: uVAR-model} to the survey data gives us optimal values of the parameters ($\beta$, $\alpha$, $\gamma$, $A_r$, $R_r$) (see Table \ref{tab:parameters}). The results described above are based on the state-level data which yields only 50 parameter values and might not capture the variability in the data. To resolve this, we perform repeated samples with replacement (known as Bootstrapping), as described in method section. The resulting distributions for the parameter means are shown in Figure~\ref{fig:VAR-Bootstrapped}. This shows that, on average, Republican states have higher values of parameters as also seen in Figure~\ref{fig:VAR_boxplotpara}. 

\begin{figure}
    \centering
    \includegraphics[scale=.4]{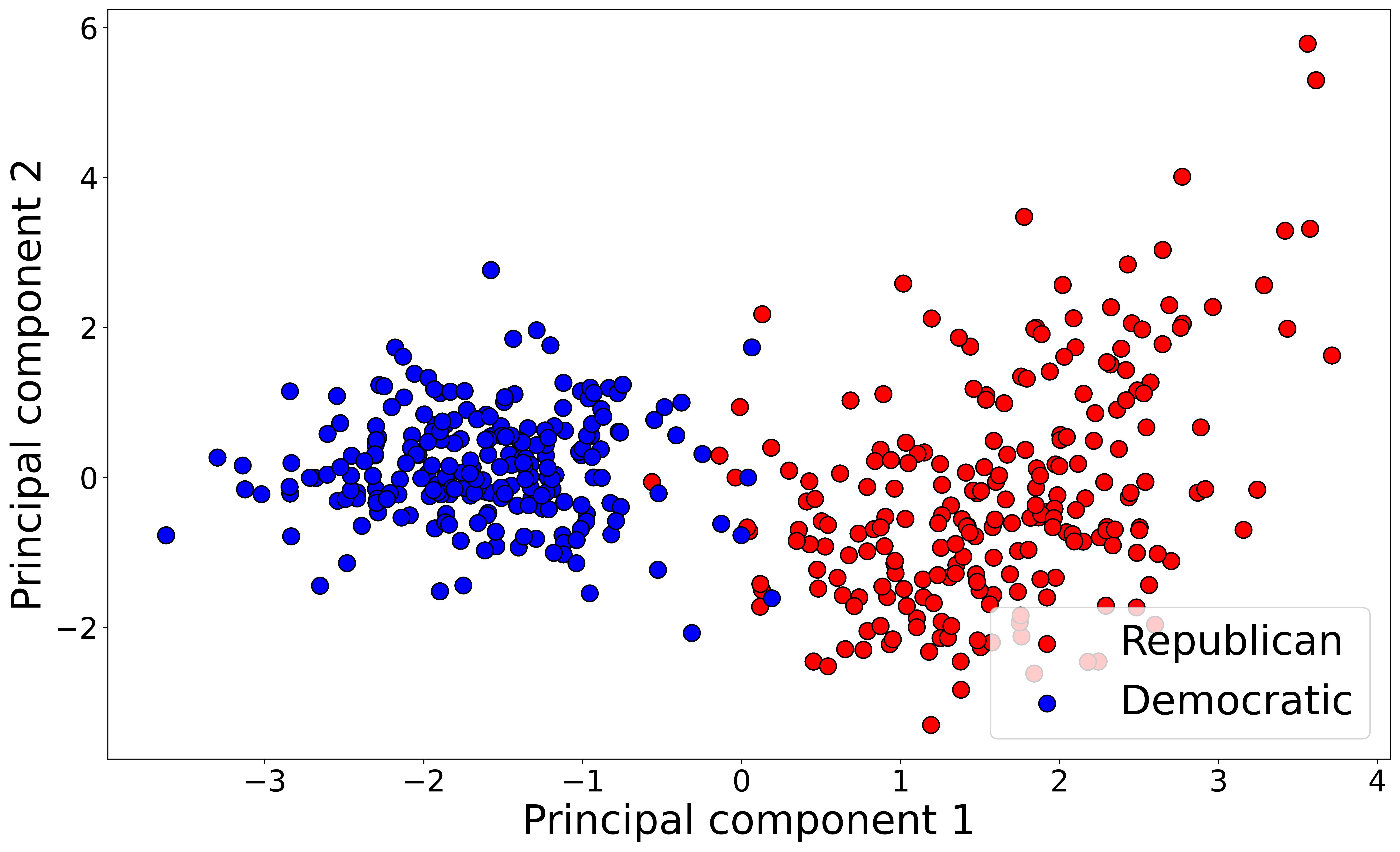}
    \caption{PCA clustering of states based on five vaccine model parameters, highlighting distinct political affiliations with Republican states in red and Democratic states in blue.}
    \label{fig:PCA_for_5_aggregated}
\end{figure}

Although the pairwise scatter plots in Figure~\ref{fig:VAR_scatter_plots} and the PCA in Figure~\ref{fig:PCA_state_wise} helped us see the clustering of the data, clustered into Democratic and Republican, there was some overlap between the two clusters (see Figure~\ref{fig:PCA_state_wise}). We used the distribution obtained from bootstrapping to see which parameter influences the data variance most. The pairwise scatter plots of parameters shown in Figure~\ref{fig:scatter_boots} are used to see which pair captures the most variance. The PCA plot (Figure~\ref{fig:PCA_for_5_aggregated}) indicates that there exist two principal components (i.e. parameters) that capture the most variance. After scaling the data, the first principal component captures $0.6296$ of the total variance, while the second component accounts for an additional $0.2707$, resulting in a cumulative explained variance of $0.9003$. This indicates that the vast majority of the variance in the dataset is preserved within the first two components, justifying the dimensionality reduction to two dimensions.

\subsection{Demographics}\label{sec:demographic_results}
Here, we explore the impact of incorporating demographic structures into our model by grouping and analysing the data based on different categories. 
The CTIS survey includes questions on age, education, ethnicity, and other demographic characteristics. Here we examine the effects of different age groups and education levels by dividing and aggregating the data according to these specific features.  
As before, we categorise all states according to their political affiliation and then aggregate the data based on demographic characteristics. 
We then optimise the sVAR Model~\eqref{equ: sVAR-model}.

\subsubsection{Age}
The survey allowed respondents to select their age from seven predefined intervals: 18-24, 25-34, 35-44, 45-54, 55-64, 65-74, and over 75.
To facilitate a more manageable analysis, we consolidate these age intervals into three broader categories: (1) Group 1: 18-34, (2) Group 2: 35-64, and (3) Group 3: above 64.
Two primary considerations guide this consolidation. 
First, it aims to achieve a relatively balanced distribution of individuals across the groups. 
Second, individuals within these broader age ranges will likely exhibit similar behavioural patterns and characteristics due to their shared life stages. 

\begin{figure}[ht]
    \centering
    
    \begin{subfigure}[b]{0.9\textwidth}
        \centering
        \includegraphics[width=\textwidth]{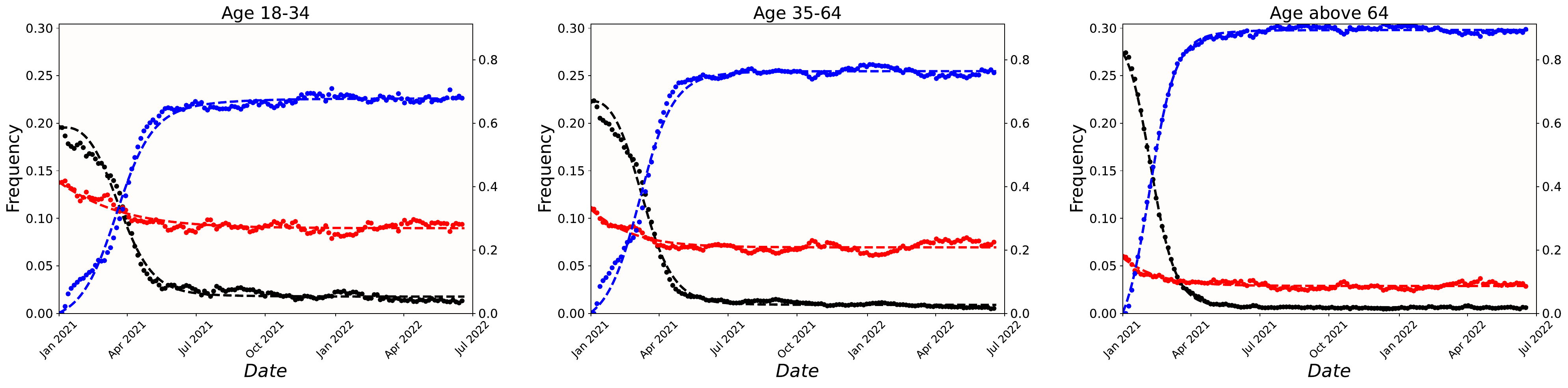}
        \caption{Republican states}
        \label{fig: age_fit_republic}
    \end{subfigure}
    \hfill
    \begin{subfigure}[b]{0.9\textwidth}
        \centering
        \includegraphics[width=\textwidth]{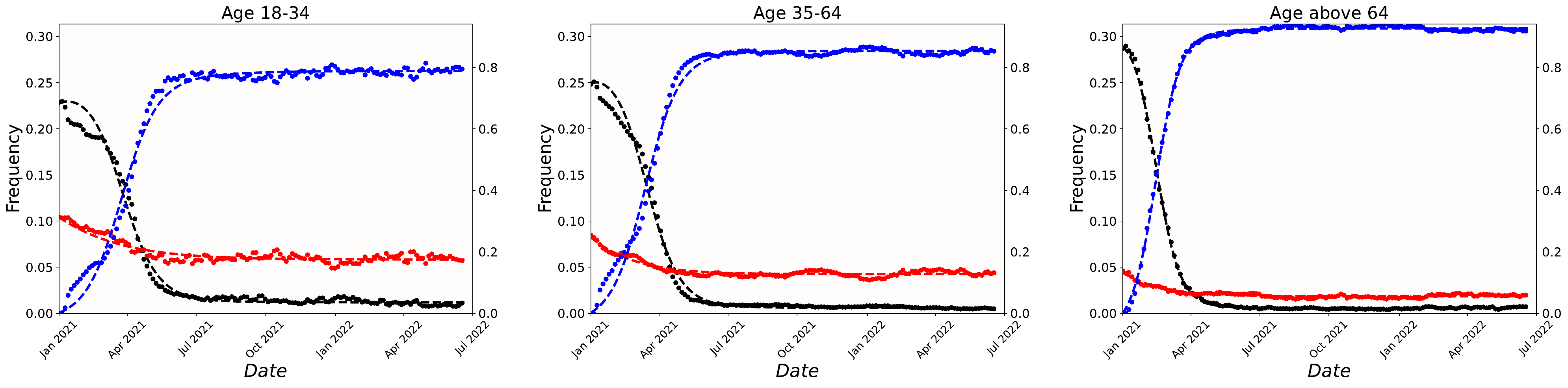}
        \caption{Democratic states}
        \label{fig: age_fit_democratic}
    \end{subfigure}
    
    \caption{Fit of the sVAR Model~\eqref{equ: sVAR-model} based on age to data from the survey where all states with the same political behaviour are aggregated together (The
solid lines are the solution curves of the sVAR Model \eqref{equ: sVAR-model}, and the dotted lines show the survey data. Blue, black, and red indicate vaccinated, accepted, and refused. The values of
parameters are shown in Figure~\ref{fig: age_params}).}
    \label{fig: age_fit}
\end{figure}

\begin{figure}
    \centering
    \includegraphics[width=0.7\linewidth]{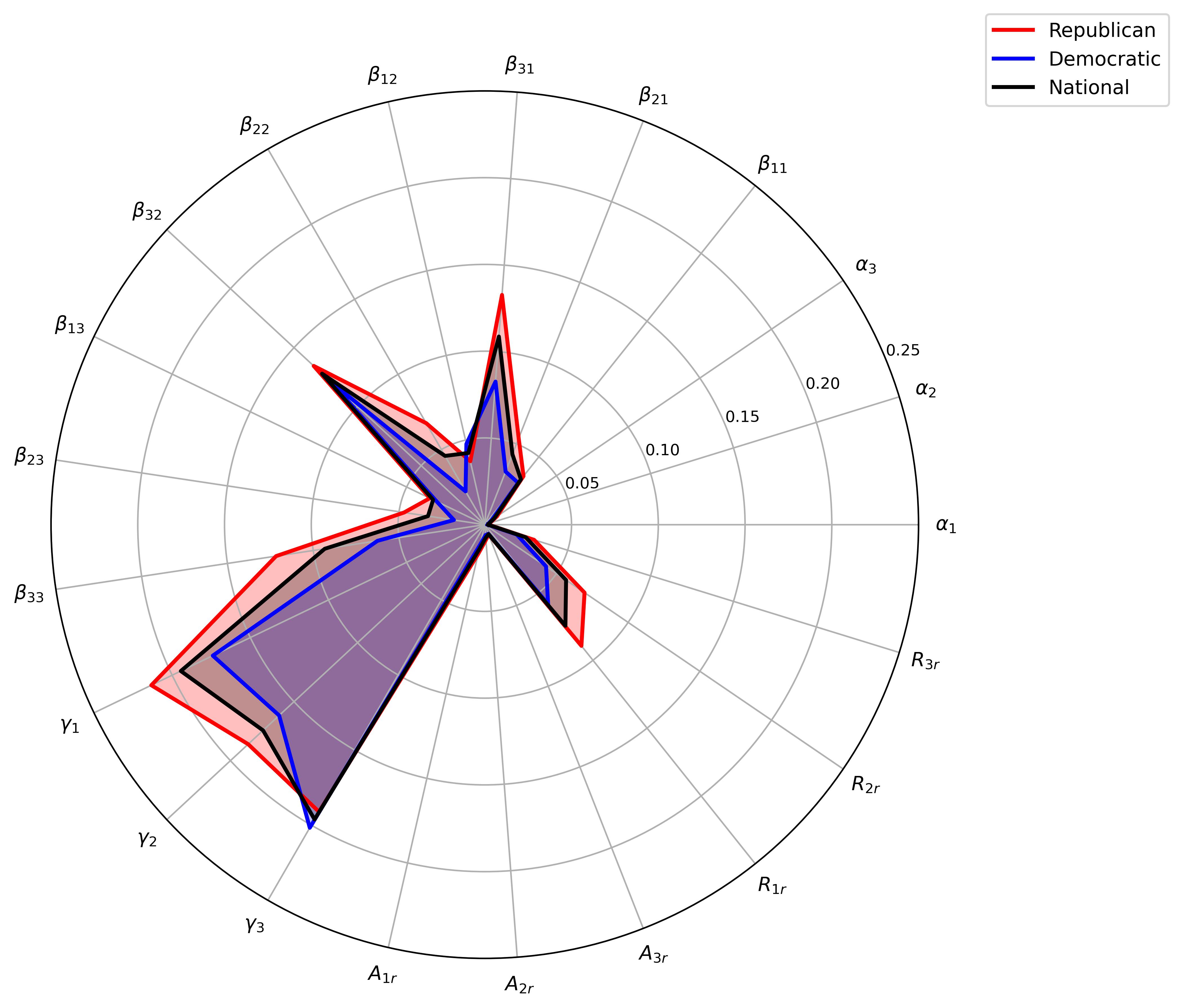}
    \caption{The average values of parameters for the sVAR Model \eqref{equ: sVAR-model} based on age such that individuals are divided into three broader categories: (1) Group 1: 18-34, (2) Group 2: 35-64, and (3) Group 3: above 64 (The single-digit indices in the parameter refer to age groups, while the two-digit indices say $ji$, which refers to the influence of age group $j$ on $i$. Also, $A_{r,i}$ and $R_{r,i}$ refer to the residuals in group $i$).}
    \label{fig: age_params}
\end{figure}

 Figure \ref{fig: age_fit} illustrates the results for Republican and Democratic states across three age classes. The data is presented using two frequency scales: the left scale depicts the frequency within the entire population, with all nine curves summing to one, while the right scale shows the frequency within each group, where the sum of Vaccinated, Refused, and Accepted frequencies equals one for each age group. Figure \ref{fig: age_params} presents the average of optimised model parameters in all states based on their political affiliation and also at the national level.

Considering the national level as the average of republic and democratic states, reveals a clear relationship between age and vaccination frequency, regardless of political affiliations. The older group consistently exhibits higher vaccination frequencies ($0.92$) compared to younger age groups ($0.73$) in their age group. Conversely, vaccine refusal rates decrease as age increases, suggesting that older individuals were more receptive to vaccination.
A pattern emerges when examining political affiliations within each age group: Democratic-leaning states generally show higher vaccination frequency. However, this political divide becomes less pronounced in older age brackets, indicating that political affiliation substantially influences younger individuals’ vaccination choices. This trend is particularly evident in Figure \ref{fig: age_params}, where comparisons of parameters such as $A_{r,i}$ and $R_{r,i}$ across political lines illustrate these differences.

The $\beta$ parameters show interesting variations across age groups and political systems. In Democratic states, the younger age group exerts a more decisive influence than the other two groups, while in the second and older age groups, the older group’s influence predominates. Conversely, in Republican states, the second group most strongly influences the first group, and the first group most strongly influences the second. Notably, the third group in Republican states exhibits high self-influence.

\begin{figure}
    \centering
    \includegraphics[scale=0.3]{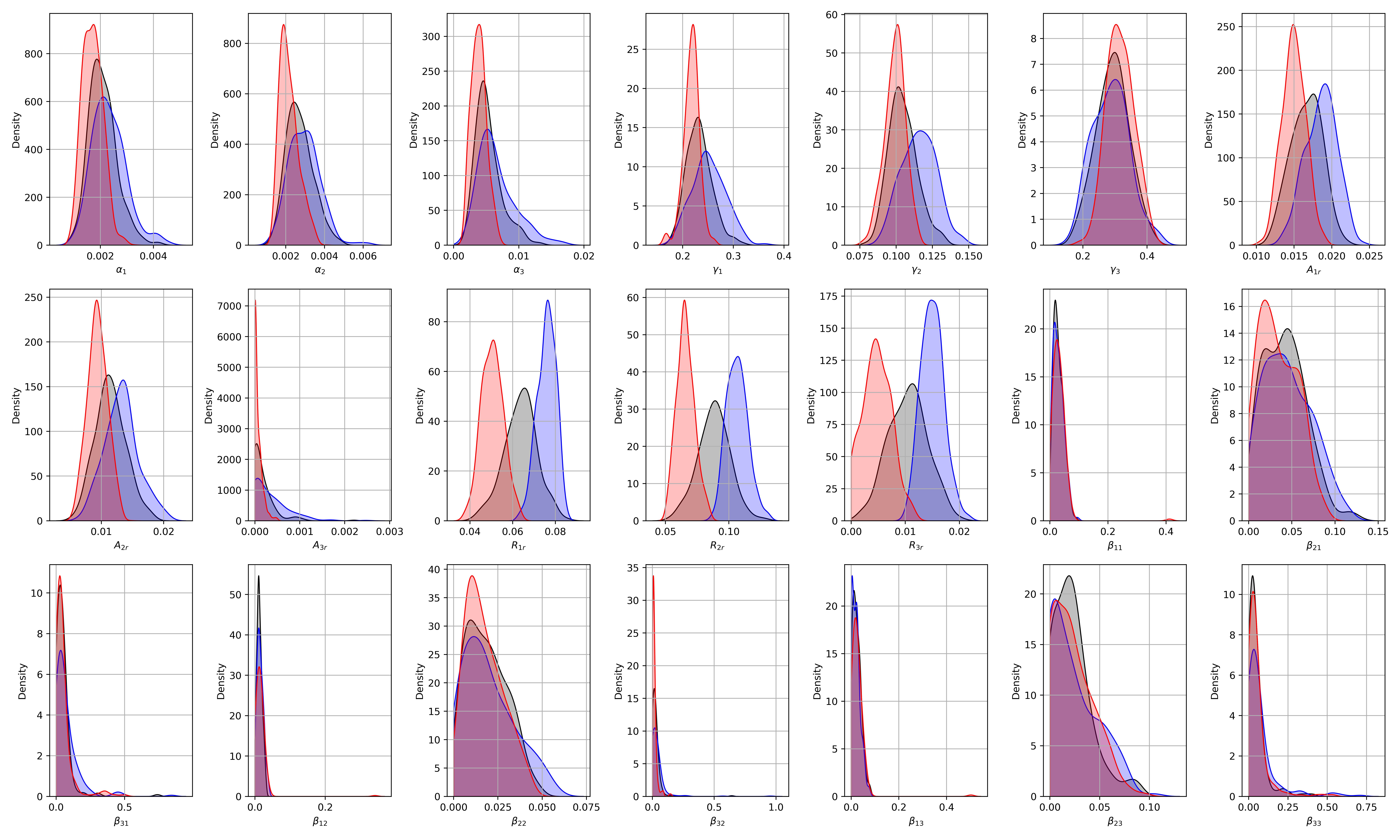}
    \caption{Distribution of parameters in sVAR Model~\eqref{equ: sVAR-model} for age using Bootstrapping method on the datasets of states separately based on political affiliations democratic (blue), republican (red) and all of them together as national (green) (In each group, we randomly selected $5$ states and aggregated them as a dataset, then we optimised our model and found parameters. We did it for each group $5000$ times).
    }
    \label{fig:age_Boostrap}
\end{figure}

To gain a deeper understanding of the behaviour, we employed the Bootstrapping method instead of relying solely on average values for each parameter. The resulting parameter distributions are presented in Figure \ref{fig:age_Boostrap}, with additional visualizations in the form of a boxplot and bar chart for the national level shown in Figures \ref{fig:age_Boostrap_boxplot} and \ref{fig:age_Boostrap_average}. These results indicate that older individuals have a stronger influence on younger individuals, consistent with the findings from the full dataset of 50 states.

\subsubsection{Education}
Here, we explore education as a key demographic factor influencing vaccination decisions. In the survey, participants were asked to select their education level from eight options: (1) Less than high school, (2) High school graduate or equivalent (GED), (3) Some college, (4) Two-year degree, (5) Four-year degree, (6) Master’s degree, (7) Professional degree (e.g., MD, JD, DVM), and (8) Doctorate. To simplify the model, we grouped these into three main categories: Group 1 includes lower education levels (options 1 and 2), Group 2 represents mid-level education (options 3, 4, and 5), and Group 3 includes advanced education (options 6, 7, and 8). These new groupings enable us to better analyse how differences in education influence vaccination decisions and opinions. Additionally, using our sVAR model, we can assess how education groups interact and influence one another.

\begin{figure}[ht]
    \centering
    
    \begin{subfigure}[b]{0.9\textwidth}
        \centering
        \includegraphics[width=\textwidth]{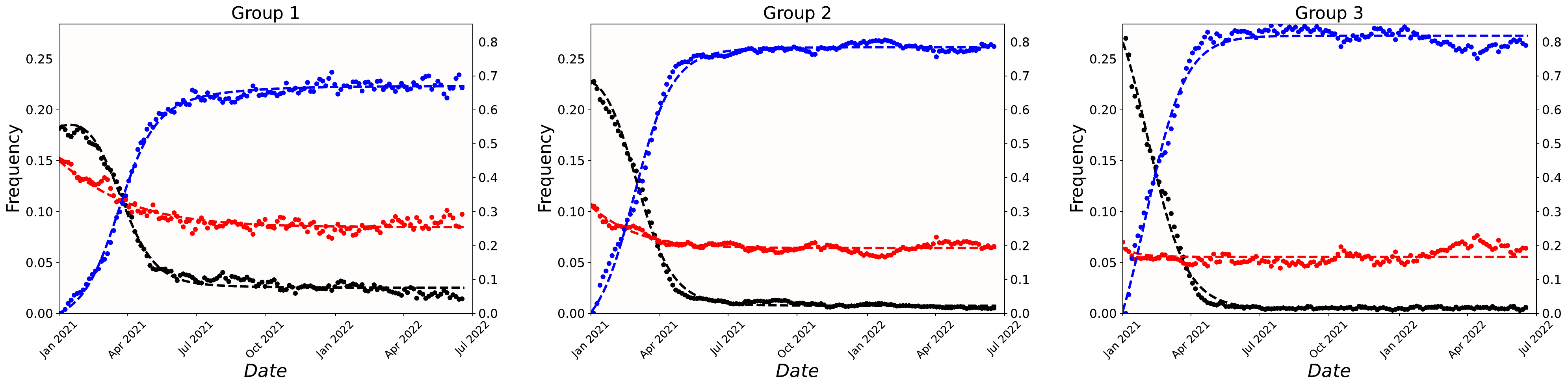}
        \caption{Republican states}
        \label{fig: edu_fit_republic}
    \end{subfigure}
    \hfill
    \begin{subfigure}[b]{0.9\textwidth}
        \centering
        \includegraphics[width=\textwidth]{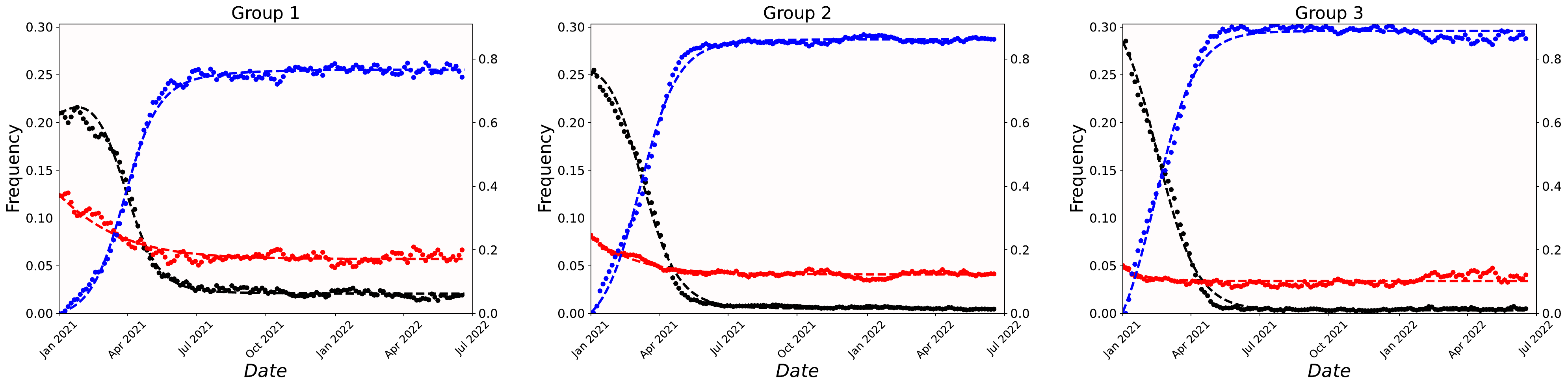}
        \caption{Democratic states}
        \label{fig: edu_fit_democratic}
    \end{subfigure}
    
    \caption{Fit of the sVAR Model~\eqref{equ: sVAR-model} based on education to data from the survey where all states with the same political behaviour are aggregated together (The
solid lines are the solution curves of the sVAR Model \eqref{equ: sVAR-model}, and the dotted lines show the survey data. Blue, black, and red indicate vaccinated, accepted, and refused. The values of
parameters are shown in Figure \ref{fig: age_params}).}
    \label{fig: edu_fit}
\end{figure}

\begin{figure}
    \centering
    \includegraphics[width=0.8\linewidth]{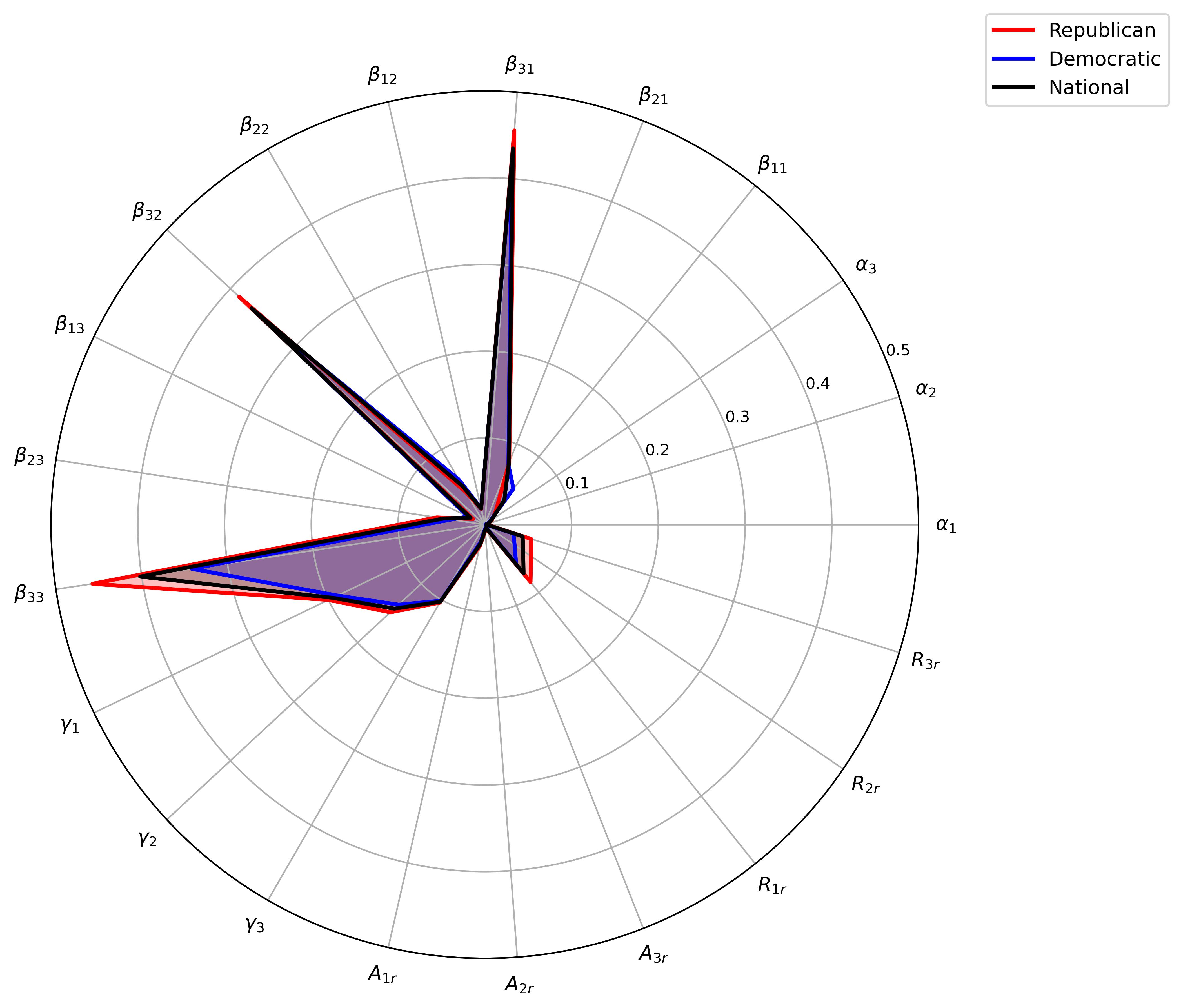}
    \caption{The average values of parameters for the sVAR Model \eqref{equ: sVAR-model} based on education such that individuals are divided into three broader categories: (1) Group 1 includes lower education levels, (2) Group 2 represents mid-level education, and (3) Group 3 includes advanced education (The single-digit indices in the parameter refer to education groups, while the two-digit indices say $ji$, which refers to the influence of age group $j$ on $i$. Also, $A_{r,i}$ and $R_{r,i}$ refer to the residuals in group $i$).}
    \label{fig: edu_params}
\end{figure}

Figure \ref{fig: edu_fit} presents the results for Republican and Democratic states across three education categories. The data is shown using two frequency scales: the left scale represents the frequency relative to the entire population, with all nine curves summing to one, while the right scale depicts the frequency within each group, where the sum of Vaccinated, Refused, and Accepted frequencies equals one for each age group. Additionally, Figure \ref{fig: edu_params} displays the average optimized model parameters across all states, categorized by political affiliation, as well as at the national level.

The data and our fitted model reveal intriguing interactions between education groups and vaccination behavior. If we consider the average of the Republican and Democratic states in Figure \ref{fig: edu_fit}, it can be seen that at the beginning of the vaccination rollout, individuals with higher education levels exhibited a greater acceptance of vaccines within their group compared to those with lower education levels, indicating that education positively influenced vaccine acceptance regardless of political affiliation. Additionally, by the end of the observation period, the parameters $A_{r,i}$ (Accepted rate) and $R_{r,i}$ (Refused rate)—which reflect the final proportions of acceptance and refusal—decrease as education level increases, signifying that highly educated individuals were more likely to be vaccinated. When examining education levels in conjunction with political affiliation as shown in Figure \ref{fig: edu_fit}, we find that, consistent with prior observations, excluding demographic factors, Democrats have a higher vaccination fraction than Republicans within each education category.

Note that the $\beta_{3i}$ values for all $i$ are consistently higher than the others. This suggests that the advanced-educated group has a stronger influence on the other education groups. In particular, it appears that individuals in the other education groups who initially refused vaccination are more likely to change their opinion due to the influence of the advanced-educated group.

\begin{figure}
    \centering
    \includegraphics[scale=0.3]{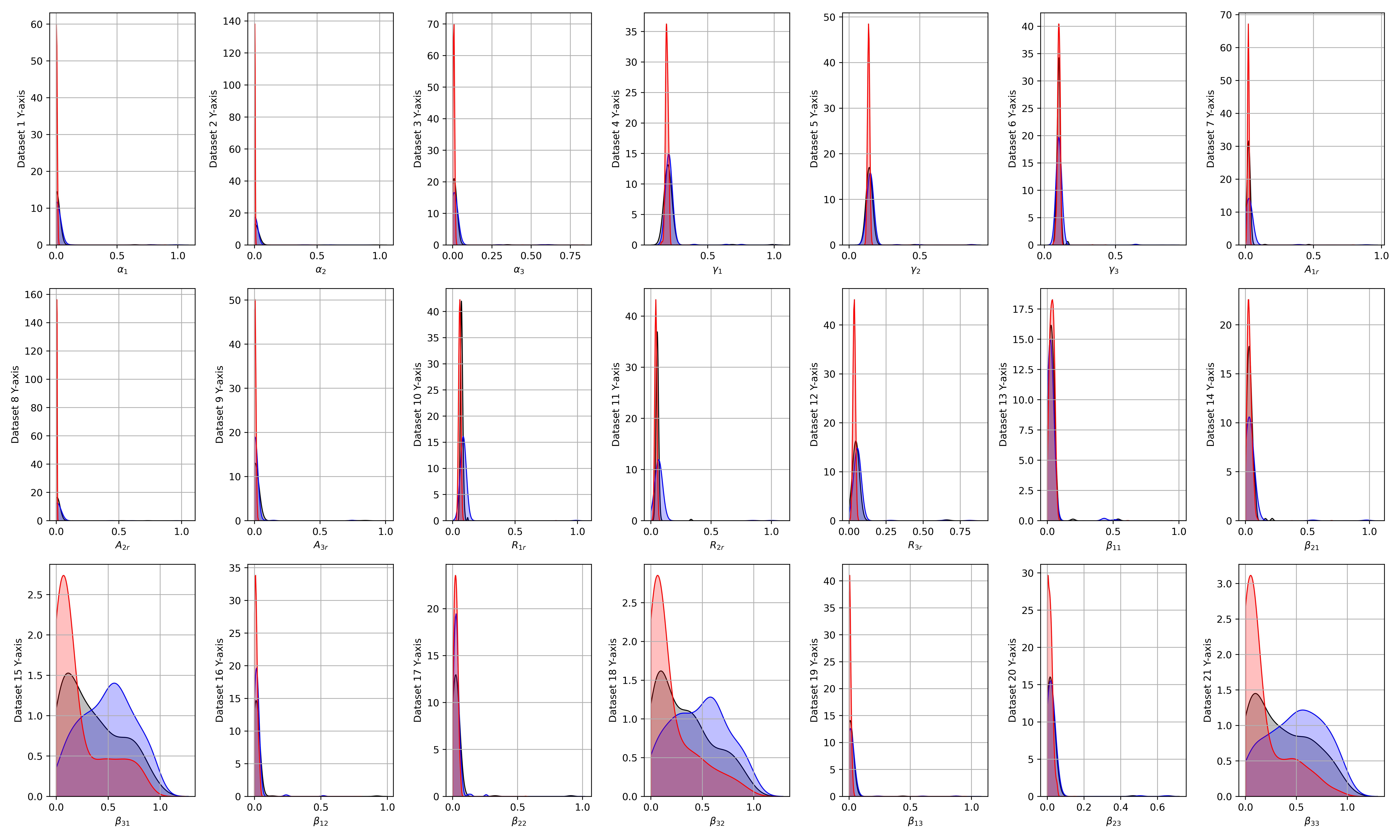}
    \caption{Distribution of parameters in sVAR Model~\eqref{equ: sVAR-model} for education using Bootstrapping method on the datasets of states separately based on political affiliations democratic (blue), republican (red) and all of them together as national (green) (In each group, we randomly selected $5$ states and aggregated them as a dataset, then we optimised our model and found parameters. We did it for each group $5000$ times). 
    }
    \label{fig:edu_Boostrap_}
\end{figure}

In this analysis, also we utilized the Bootstrapping method. The resulting parameter distributions are shown in Figure \ref{fig:edu_Boostrap_}, with supplementary visualizations—including a boxplot and bar chart for the national level—provided in Figures \ref{fig:edu_Boostrap_box} and \ref{fig:edu_Boostrap_average}. These results suggest that individuals with higher levels of education exert a stronger influence on those with lower education levels, aligning with findings from the full dataset encompassing all states.

\section{Discussion}\label{sec:discussion}
Vaccination is a widely used mechanism to help protect people from infectious diseases and to reduce their spread~\cite{plotkin2014history}. 
At a population level, high vaccination rates are often  required to achieve herd immunity~\cite{koopman2017dynamics}. 
However, people often need more motivation to be vaccinated, as they often show hesitancy and delay or even refuse it~\cite{Amin2017,barello2020vaccine,Hammershaimb2023}. 
Understanding public attitudes towards vaccination is crucial for estimating the success of vaccination campaigns against  infectious diseases and highly prevalent non-communicable diseases. 
Social interaction is one of the key drivers of diverse responses to vaccination~\cite{cascini2021attitudes}. 
Mathematical models allow us to quantify these interactions and identify the population's most and least influential demographic groups.

We have utilized data from the COVID-19 Trends and Impact Survey (CTIS), conducted by the Delphi Research Group at Carnegie Mellon University in collaboration with Facebook~\cite{Salomon2021}. This large-scale survey, with daily responses  across the USA has provided detailed insights into vaccination status, attitudes, and demographics, enabling analysis of vaccine acceptance and hesitancy at national and state levels\cite{bui2025analysis,sasse2024understanding,soorapanth2023rural,ronn2023vaccination,henneberger2022estimates}. Using this dataset, we have divided the population into three classes—Vaccinated, Accepted, and Refused—and have developed mechanistic compartmental models to describe transitions between these groups over time, governed by mass-action dynamics. We have examined how political affiliations, based on the 2020 USA presidential election, have influenced vaccination behaviour, analysing dynamics with and without demographic structures. Our findings show that model parameters effectively distinguish states by political affiliation, revealing differing COVID-19 vaccine responses among Democrats and Republicans. Using the simplified Vaccinated-Accepted-Refused (uVAR) model, we confirm and extend previous findings on political affiliation’s impact on vaccine uptake \cite{albrecht2022vaccination, khubchandani2021covid, ronn2023vaccination}. Our analysis across USA states quantitatively supports that political attitudes shape vaccination behaviour, with Democratic-leaning states exhibiting higher final vaccinated proportions and lower transition rates, reflecting stronger initial acceptance. Principal Component Analysis (PCA) further validates these results, showing clear clustering by political affiliation, reinforcing the link between trust in government and vaccine hesitancy observed in earlier studies.

Vaccine hesitancy can be influenced by several demographic factors, including age, education, ethnic, and gender~\cite{malik2020determinants,lazarus2023hesitant}. Among these, we have focused on age and education to investigate their impact on vaccine hesitancy using our sVAR model, which has been designed to capture interactions within and between demographic groups. The population has been divided into three age groups (18–24, 35–64, 65+) and three education levels (high school or less, some college, four-year degree or higher) to analyse how these factors shape public opinion on COVID-19 vaccination. Our two-step approach, combining demographic groupings with political affiliations, has confirmed also that both age and education can differently impact vaccination behaviour.

The age-structured analysis showed that baseline vaccination rates increased with age, as older groups were prioritized when vaccines first became available. Although vaccination progress over time was similar across age groups, the youngest cohort (18–24) displayed the highest vaccine hesitancy, indicated by larger residuals of unvaccinated individuals. In contrast, older age groups (65+) were the least hesitant. Positive social influence on vaccine acceptance also strengthened with age, suggesting older individuals were more effective in encouraging others to vaccinate. On average republican-leaning states exhibited higher hesitancy rates across age groups compared to Democratic states, largely due to lower initial acceptance levels. Interestingly, in the older age group, both Republican and Democratic states had similar vaccination rates, likely because older individuals prioritized their health over political factors. However, in the younger group, vaccination rates were higher in Democratic states, possibly reflecting that younger individuals are more influenced by government policies and political affiliations when making health decisions. In the education-structured analysis, baseline vaccination rates were consistent across all education levels, reflecting equal vaccine access regardless of educational background. However, hesitancy was higher among individuals with lower education levels (high school or less), as seen in the larger residuals of unvaccinated individuals. Those with advanced degrees showed the lowest hesitancy and a stronger influence on promoting vaccine acceptance within their communities. These results highlight that both age and education significantly affect vaccine hesitancy, with older and more educated individuals being more likely to accept vaccination and positively influence others.

In summary, our approach introduced a novel mathematical model to quantify vaccine hesitancy dynamics, incorporating political and demographic influences to understand how attitudes toward COVID-19 vaccination evolve over time. By structuring the population into vaccinated, accepted, and refused classes, our model captures transitions driven by social interactions and external factors. Applying this model to USA survey data, we quantify the impact of political affiliation, age, and education on vaccine uptake and hesitancy, revealing significant ideological and demographic disparities. Our findings offer a data-driven framework and mathematical modelling for tailoring public health strategies and improving vaccine adoption through targeted interventions.

While our models provide valuable insights, they simplify complex human behaviours by assuming homogeneity within demographic groups and constant transition rates. In reality, individual decisions evolve over time, influenced by shifting social, political, and personal factors. Future work should focus on developing models that capture individual-level dynamics, tracking changes in vaccination attitudes over time. Longitudinal survey data, following the same individuals across multiple time points, would be essential for this deeper analysis. Additionally, while we focused on age and education, other demographic factors like gender, ethnic, and socioeconomic status could further illuminate the drivers of vaccine hesitancy. Expanding the model to incorporate multiple demographic layers would provide a more comprehensive understanding of the issue. Moreover, applying our framework to data from other countries would allow for cross-cultural comparisons, helping to identify universal and context-specific drivers of vaccine hesitancy. Finally, while our study used survey data as a proxy for population behavior, more granular data, including individual-level responses, would enhance the accuracy and predictive power of the models. Such improvements could inform more nuanced public health interventions, ultimately contributing to higher vaccination rates and better health outcomes.

\section*{Acknowledgements}
This project has received funding from the European Union’s Horizon 2020 research and innovation programme under grant agreement No 955708.

\section*{Ethical Statement}
The Ethics Board (IRB) of IMDEA Networks Institute approved this work on 2021/07/05. IMDEA Networks has signed Data Use Agreements with Facebook and Carnegie Mellon University (CMU) to access their data. Specifically, CMU project STUDY2020\_00000162 entitled ILI Community-Surveillance Study. Informed consent has been obtained from all participants in this survey by this institution. All the methods in this study have been carried out following relevant ethics and privacy guidelines and regulations.

\section*{Availability of Data}
The microdata of the CTIS survey from which the aggregated data was obtained cannot be shared, as per the Data Use Agreements signed with Facebook and Carnegie Mellon University (CMU). The aggregated data and code supporting the findings of this study are available at \url{https://tinyurl.com/wf5abc43}.

\bibliographystyle{abbrv}  
\bibliography{library}
\makeatletter
\renewcommand \thesection{S\@arabic\c@section}
\renewcommand\thetable{S\@arabic\c@table}
\renewcommand \thefigure{S\@arabic\c@figure}
\makeatother
\renewcommand{\theequation}{S.\arabic{equation}}
\setcounter{equation}{0}
\renewcommand{\thetable}{S.\arabic{table}}  
\setcounter{table}{0}  %
\renewcommand{\thefigure}{S.\arabic{figure}}  
\setcounter{figure}{0} 

\section{Supplementary Information}
\subsection{Parameter Optimization}\label{sec: optimisation}
We can choose from several methods available to estimate the unknown parameters by parameters by fitting models to the survey data described in Section~\ref{sec:materials_methods}. We choose to estimate the parameters by using used least squares. 
In the next section we describe the optimization scheme that we have used to minimize our functions. 
\subsubsection{Differential Evolution (DE)}\label{sec: differential evolution}
The Differential Evolution (DE) method is an iterative metaheuristic method that makes very few assumptions for searching an optimal solution in a large parameter space. That is, the scheme generate $m$-dimensional trial space of parameter vectors ($\lambda$) and $f(\lambda)$ is obtain for their fitness. Then for every vector ($\lambda_i$) in the trial space, a new point (say $\lambda_z$) is constructed from randomly chosen three vectors (say $\lambda_a,\lambda_b,\lambda_c$) by adding the weighted difference of the two vectors ($w(\lambda_b-\lambda_c)$) to the third vector ($\lambda_a$). The function $f$ will be evaluated at this newly constructed vector $\lambda_z$ and the fitness is compared with original vector $\lambda_i$ and will replace it if $f(\lambda_z)<f(\lambda_i)$.

\subsection{Principal Component Analysis (PCA)}\label{sec:PCA}

PCA is a dimensionality reduction technique that identifies the directions in the feature space, called principal components, which capture the maximum variance in the data. To achieve this, we first standardize the dataset \( X \), consisting of \( m \) samples and \( n \) features, as follows:
\begin{equation}
X_{\text{std}} = \frac{X - \mu}{\sigma},
\end{equation}
where \( \mu \) is the mean vector and \( \sigma \) is the standard deviation vector of \( X \).

Next, we compute the covariance matrix of the standardized data:
\begin{equation}
\Sigma = \frac{1}{m-1} X_{\text{std}}^\top X_{\text{std}},
\end{equation}
where \( \Sigma \) is an \( n \times n \) symmetric matrix. The principal components are then obtained by solving the eigenvalue problem:
\begin{equation}
\Sigma \mathbf{v}_i = \lambda_i \mathbf{v}_i,
\end{equation}
where \( \lambda_i \) are the eigenvalues, and \( \mathbf{v}_i \) are the corresponding eigenvectors. The eigenvectors \( \mathbf{v}_1 \) and \( \mathbf{v}_2 \) associated with the two largest eigenvalues \( \lambda_1 \) and \( \lambda_2 \) define the first two principal components.

Finally, the data is projected onto the subspace spanned by these components using:
\begin{equation}
Z = X_{\text{std}} \cdot V,
\end{equation}
where \( V = [\mathbf{v}_1, \mathbf{v}_2] \) is the matrix of the top two eigenvectors, and \( Z \) is the transformed data in the two-dimensional PCA space. This projection reduces the dimensionality of the data while preserving the majority of its variance, enabling simplified analysis and visualization.

\subsection{Mann–Whitney U Test}\label{sec:MWU-test}

The Mann–Whitney U test is a non-parametric test used to compare two independent groups, particularly when the data are not normally distributed. Unlike parametric tests that compare means, the Mann–Whitney U test evaluates whether the ranks of one group tend to be higher or lower than those of the other group. The null hypothesis for the test states that the distributions of the two groups are the same, while the alternative hypothesis suggests that there is a difference between the distributions. The $p$-value derived from the test indicates the statistical significance of the difference. A p-value less than 0.05 typically suggests that the two groups differ significantly.

The test works by first combining the data from both groups and ranking all the values together. Then, the sum of ranks for each group is computed, and the U statistic is calculated. The U statistic is compared to critical values from a Mann–Whitney U table or used to derive the p-value with statistical software. One of the main advantages of the Mann–Whitney U test is that it does not assume normality, making it suitable for small sample sizes and data with outliers. It is widely used for comparing ordinal data or non-normally distributed continuous data, providing a robust and simple alternative to parametric tests.

\subsection{Additional Results}
\begin{figure}[th]
    \centering
    \includegraphics[scale=.35]{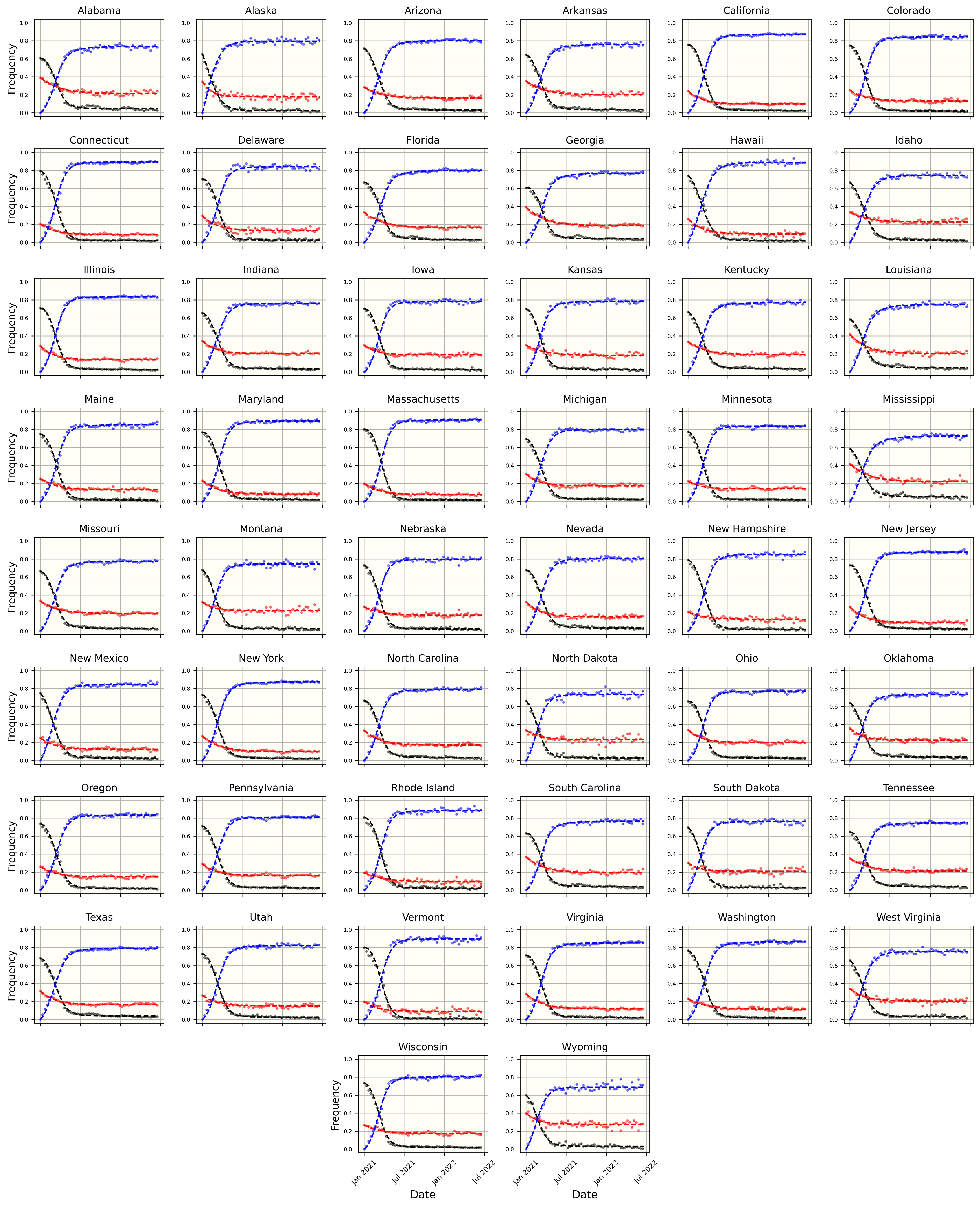}
    \caption{The prediction of the sVAR Model~\eqref{equ: sVAR-model} and data from the survey for all the states (The lines show the results from model \eqref{equ: uVAR-model} and the dots show the survey data. Blue demonstrates the frequency of vaccinated people, green shows the frequency of people who are not vaccinated but accepted to be vaccinated, and red is for those who refused the vaccine.  The parameters of $\{\beta$, $\alpha$ , $\gamma\}$ are shown in Table 1).  }
    \label{fig:fitVAR_all_states}
\end{figure}

\begin{figure}[th]
    \centering
    \includegraphics[scale=.5]{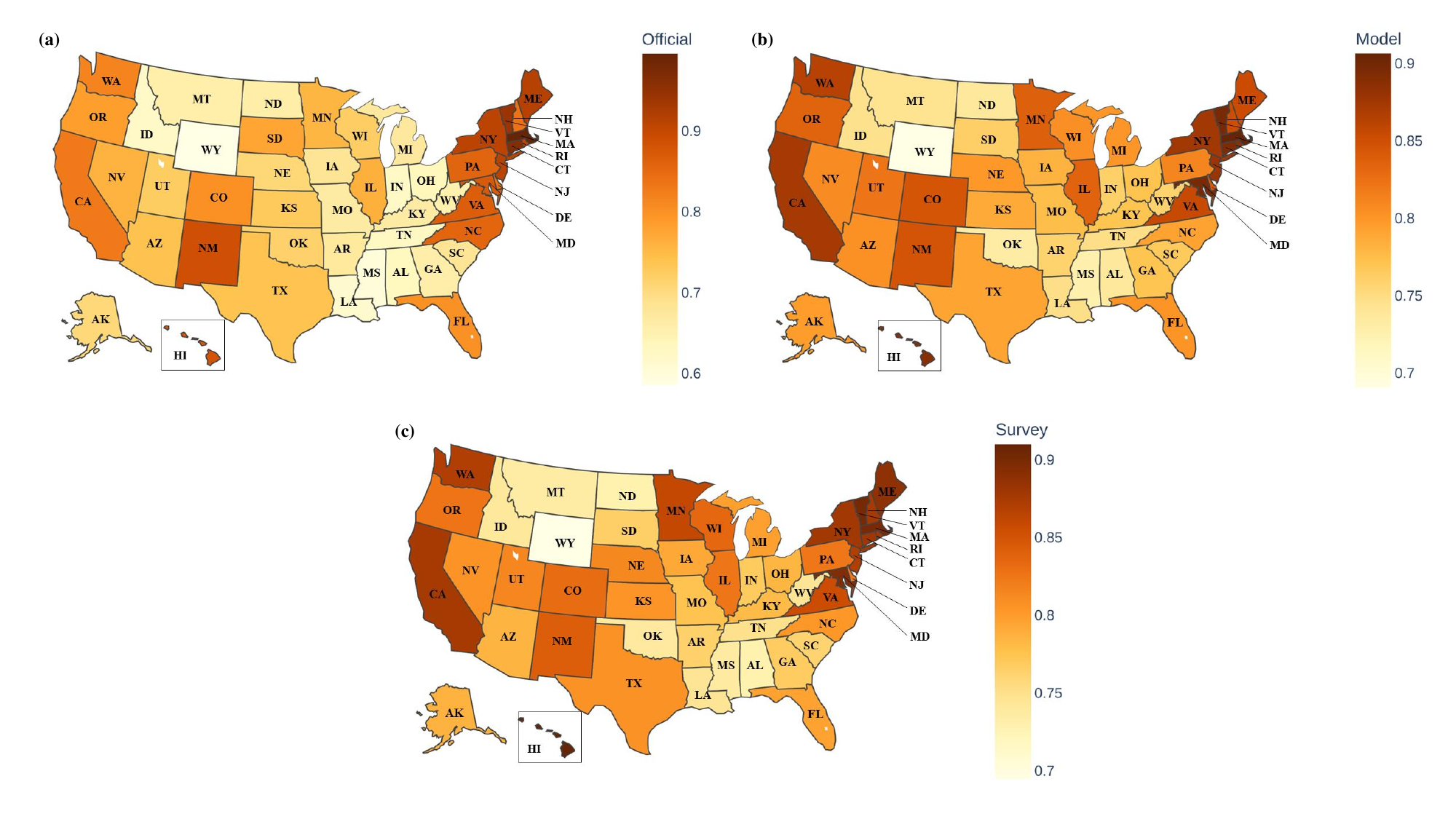}
    \caption{Final frequency of vaccinated people in (a) official data, (b) analytical solution of the uVAR Model~\eqref{equ: uVAR-model}, and (c) survey data (For the analytical solution, the parameter and initial values are given in Table \ref{tab:parameters} and Table \ref{tab2}). }
    \label{fig:Vaccination_survery_model_official}
\end{figure}

\begin{table}
    \centering
    \caption{The optimized parameter for the uVAR~\eqref{equ: uVAR-model}, and State Information}
    \begin{tabular}{llllllll}
        \toprule
        State & State Code & Politic & $\beta$ & $\alpha$ & $\gamma$ & $A_b$ & $R_b$ \\
        \midrule
        Alabama & AL & R & 0.018 & 0.004 & 0.057 & 0.047 & 0.215 \\
        Alaska & AK & R & 0.033 & 0.014 & 0.028 & 0.025 & 0.178 \\
        Arizona & AZ & D & 0.017 & 0.004 & 0.046 & 0.032 & 0.164 \\
        Arkansas & AR & R & 0.019 & 0.005 & 0.051 & 0.036 & 0.205 \\
        California & CA & D & 0.020 & 0.003 & 0.049 & 0.028 & 0.099 \\
        Colorado & CO & D & 0.019 & 0.003 & 0.048 & 0.021 & 0.133 \\
        Connecticut & CT & D & 0.019 & 0.003 & 0.043 & 0.019 & 0.087 \\
        Delaware & DE & D & 0.021 & 0.003 & 0.051 & 0.025 & 0.133 \\
        Florida & FL & R & 0.017 & 0.003 & 0.050 & 0.034 & 0.164 \\
        Georgia & GA & D & 0.018 & 0.003 & 0.058 & 0.042 & 0.188 \\
        Hawaii & HI & D & 0.019 & 0.006 & 0.035 & 0.019 & 0.093 \\
        Idaho & ID & R & 0.019 & 0.005 & 0.044 & 0.025 & 0.229 \\
        Illinois & IL & D & 0.023 & 0.003 & 0.049 & 0.027 & 0.138 \\
        Indiana & IN & R & 0.024 & 0.004 & 0.046 & 0.033 & 0.205 \\
        Iowa & IA & R & 0.024 & 0.003 & 0.056 & 0.027 & 0.190 \\
        Kansas & KS & R & 0.016 & 0.003 & 0.056 & 0.029 & 0.184 \\
        Kentucky & KY & R & 0.018 & 0.004 & 0.052 & 0.037 & 0.192 \\
        Louisiana & LA & R & 0.020 & 0.006 & 0.052 & 0.046 & 0.206 \\
        Maine & ME & D & 0.017 & 0.003 & 0.048 & 0.017 & 0.130 \\
        Maryland & MD & D & 0.017 & 0.002 & 0.047 & 0.021 & 0.083 \\
        Massachusetts & MA & D & 0.017 & 0.002 & 0.048 & 0.016 & 0.077 \\
        Michigan & MI & D & 0.024 & 0.004 & 0.044 & 0.025 & 0.175 \\
        Minnesota & MN & D & 0.023 & 0.003 & 0.049 & 0.020 & 0.143 \\
        Mississippi & MS & R & 0.018 & 0.007 & 0.051 & 0.052 & 0.221 \\
        Missouri & MO & R & 0.019 & 0.004 & 0.055 & 0.029 & 0.196 \\
        Montana & MT & R & 0.030 & 0.005 & 0.052 & 0.026 & 0.229 \\
        Nebraska & NE & R & 0.020 & 0.003 & 0.053 & 0.024 & 0.178 \\
        Nevada & NV & D & 0.021 & 0.004 & 0.047 & 0.035 & 0.157 \\
        New Hampshire & NH & D & 0.017 & 0.003 & 0.046 & 0.019 & 0.130 \\
        New Jersey & NJ & D & 0.021 & 0.004 & 0.042 & 0.025 & 0.097 \\
        New Mexico & NM & D & 0.020 & 0.006 & 0.040 & 0.028 & 0.125 \\
        New York & NY & D & 0.016 & 0.004 & 0.045 & 0.025 & 0.100 \\
        North Carolina & NC & R & 0.020 & 0.004 & 0.055 & 0.034 & 0.173 \\
        \bottomrule
    \end{tabular}
    \label{tab:parameters}
\end{table}

\begin{table}
    \centering
    \begin{tabular}{llllllll}
        \toprule
        State & State Code & Politic & $\beta$ & $\alpha$ & $\gamma$ & $A_b$ & $R_b$ \\
        \midrule
        North Dakota & ND & R & 0.022 & 0.006 & 0.052 & 0.030 & 0.232 \\
        Ohio & OH & R & 0.023 & 0.003 & 0.060 & 0.029 & 0.200 \\
        Oklahoma & OK & R & 0.027 & 0.006 & 0.050 & 0.041 & 0.227 \\
        Oregon & OR & R & 0.019 & 0.003 & 0.043 & 0.017 & 0.147 \\
        Pennsylvania & PA & D & 0.021 & 0.003 & 0.047 & 0.025 & 0.163 \\
        Rhode Island & RI & D & 0.013 & 0.002 & 0.050 & 0.021 & 0.092 \\
        South Carolina & SC & R & 0.019 & 0.004 & 0.057 & 0.039 & 0.195 \\
        South Dakota & SD & R & 0.028 & 0.005 & 0.048 & 0.029 & 0.209 \\
        Tennessee & TN & R & 0.018 & 0.003 & 0.059 & 0.040 & 0.212 \\
        Texas & TX & R & 0.021 & 0.004 & 0.044 & 0.041 & 0.167 \\
        Utah & UT & R & 0.019 & 0.003 & 0.048 & 0.027 & 0.151 \\
        Vermont & VT & D & 0.021 & 0.002 & 0.045 & 0.010 & 0.093 \\
        Virginia & VA & D & 0.018 & 0.003 & 0.049 & 0.025 & 0.121 \\
        Washington & WA & D & 0.016 & 0.003 & 0.044 & 0.019 & 0.117 \\
        West Virginia & WV & R & 0.022 & 0.006 & 0.045 & 0.034 & 0.207 \\
        Wisconsin & WI & D & 0.018 & 0.003 & 0.053 & 0.021 & 0.175 \\
        Wyoming & WY & R & 0.029 & 0.007 & 0.053 & 0.032 & 0.277 \\
        \bottomrule
    \end{tabular}
    \label{tab:parameters_page2}
\end{table}

\begin{table}
    \centering
    \caption{Final frequency of vaccinated people from official dataset, survey and model by 20th July 2022 and the initial frequency of accepted and refused people by 20th January 2021. }
    \begin{tabular}{l|ccccc||l|ccccc}
        \hline
        State & Real&Survey& Model&A(0)& R(0)&State&Real&Survey &Model&A(0)& R(0) \\
        \hline
        AL &0.633 & 0.730 & 0.738 & 0.609 & 0.391 & MT &0.658 & 0.736 & 0.744 & 0.679 & 0.321 \\
        AK &0.706 & 0.787 & 0.796 & 0.652 & 0.348 & NE &0.708 & 0.814 & 0.798 & 0.731 & 0.269 \\
        AZ &0.741 & 0.785 & 0.803 & 0.715 & 0.285 & NV &0.760 & 0.806 & 0.807 & 0.677 & 0.323 \\
        AR &0.674 & 0.764 & 0.759 & 0.646 & 0.354 & NH &0.840 & 0.876 & 0.851 & 0.790 & 0.210 \\
        CA &0.830 & 0.876 & 0.872 & 0.757 & 0.243 & NJ &0.912 & 0.870 & 0.878 & 0.732 & 0.268 \\
        CO &0.800 & 0.832 & 0.845 & 0.749 & 0.252 & NM &0.889 & 0.842 & 0.846 & 0.748 & 0.252 \\
        CT &0.966 & 0.880 & 0.894 & 0.794 & 0.206 & NY &0.909 & 0.878 & 0.874 & 0.729 & 0.271 \\     
        DE &0.840 & 0.824 & 0.841 & 0.702 & 0.298 & NC &0.856 & 0.803 & 0.792 & 0.665 & 0.335 \\
        FL &0.800 & 0.797 & 0.801 & 0.666 & 0.334 & ND &0.661 & 0.730 & 0.737 & 0.663 & 0.337 \\
        GA &0.660 & 0.768 & 0.770 & 0.608 & 0.392 & OH &0.636 & 0.784 & 0.771 & 0.659 & 0.341 \\
        HI &0.881 & 0.910 & 0.888 & 0.741 & 0.259 & OK &0.717 & 0.740 & 0.732 & 0.640 & 0.360 \\
        ID &0.622 & 0.742 & 0.745 & 0.665 & 0.335 & OR &0.785 & 0.826 & 0.835 & 0.739 & 0.261 \\ 
        IL &0.763 & 0.826 & 0.834 & 0.709 & 0.291 & PA &0.859 & 0.824 & 0.811 & 0.709 & 0.291 \\
        IN &0.625 & 0.769 & 0.761 & 0.655 & 0.345 & RI &1.001 & 0.877 & 0.886 & 0.807 & 0.193 \\
        IA &0.684 & 0.792 & 0.782 & 0.703 & 0.297 & SC &0.683 & 0.762 & 0.765 & 0.632 & 0.368 \\
        KS &0.730 & 0.805 & 0.787 & 0.702 & 0.298 & SD &0.776 & 0.764 & 0.761 & 0.696 & 0.304 \\
        KY &0.666 & 0.781 & 0.770 & 0.666 & 0.334 & TN &0.627 & 0.751 & 0.748 & 0.646 & 0.354 \\
        LA &0.615 & 0.746 & 0.747 & 0.583 & 0.417 & TX &0.739 & 0.807 & 0.791 & 0.683 & 0.317 \\
        ME &0.916 & 0.889 & 0.852 & 0.748 & 0.252 & UT &0.725 & 0.815 & 0.821 & 0.731 & 0.269 \\
        MD &0.873 & 0.895 & 0.896 & 0.768 & 0.232 & VT &0.944 & 0.902 & 0.897 & 0.800 & 0.200 \\
        MA &0.995 & 0.901 & 0.906 & 0.802 & 0.198 & VA &0.868 & 0.856 & 0.854 & 0.714 & 0.286 \\
        MI &0.674 & 0.798 & 0.800 & 0.695 & 0.305 & WA &0.817 & 0.870 & 0.863 & 0.767 & 0.233 \\
        MN &0.756 & 0.861 & 0.837 & 0.775 & 0.225 & WV &0.655 & 0.746 & 0.759 & 0.658 & 0.342 \\
        MS &0.601 & 0.739 & 0.726 & 0.585 & 0.415 & WI &0.723 & 0.835 & 0.804 & 0.733 & 0.267 \\
        MO &0.667 & 0.776 & 0.775 & 0.664 & 0.336 & WY &0.585 & 0.695 & 0.690 & 0.600 & 0.400 \\
    \end{tabular}
    \label{tab2}
\end{table}

\begin{figure}
    \centering
    \includegraphics[scale=0.3]{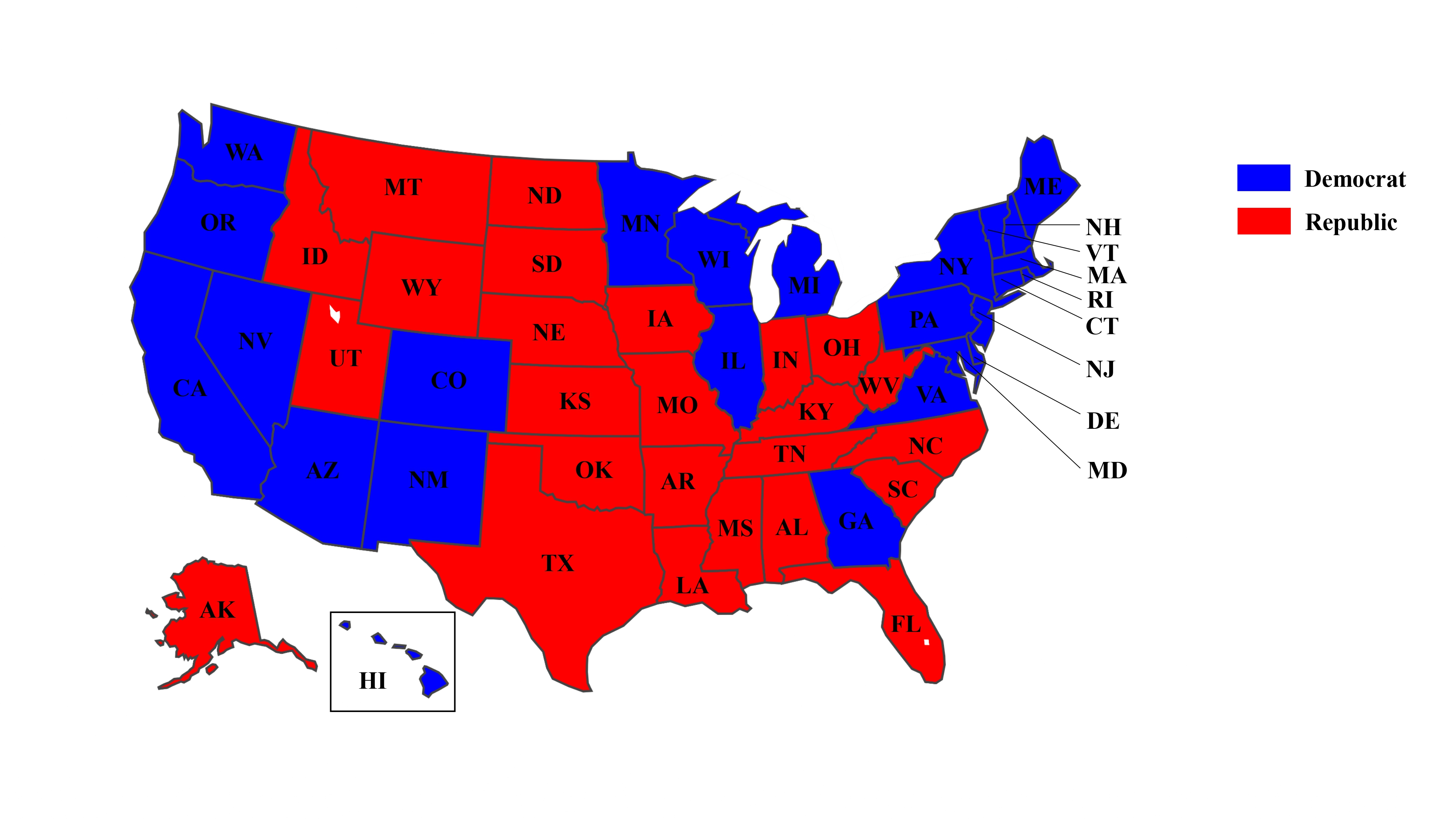}
    \caption{Political affiliation based on 2020 presidential election results.}
    \label{fig:political_affiliation}
\end{figure}

\begin{figure}
    \centering
    \includegraphics[scale=0.38]{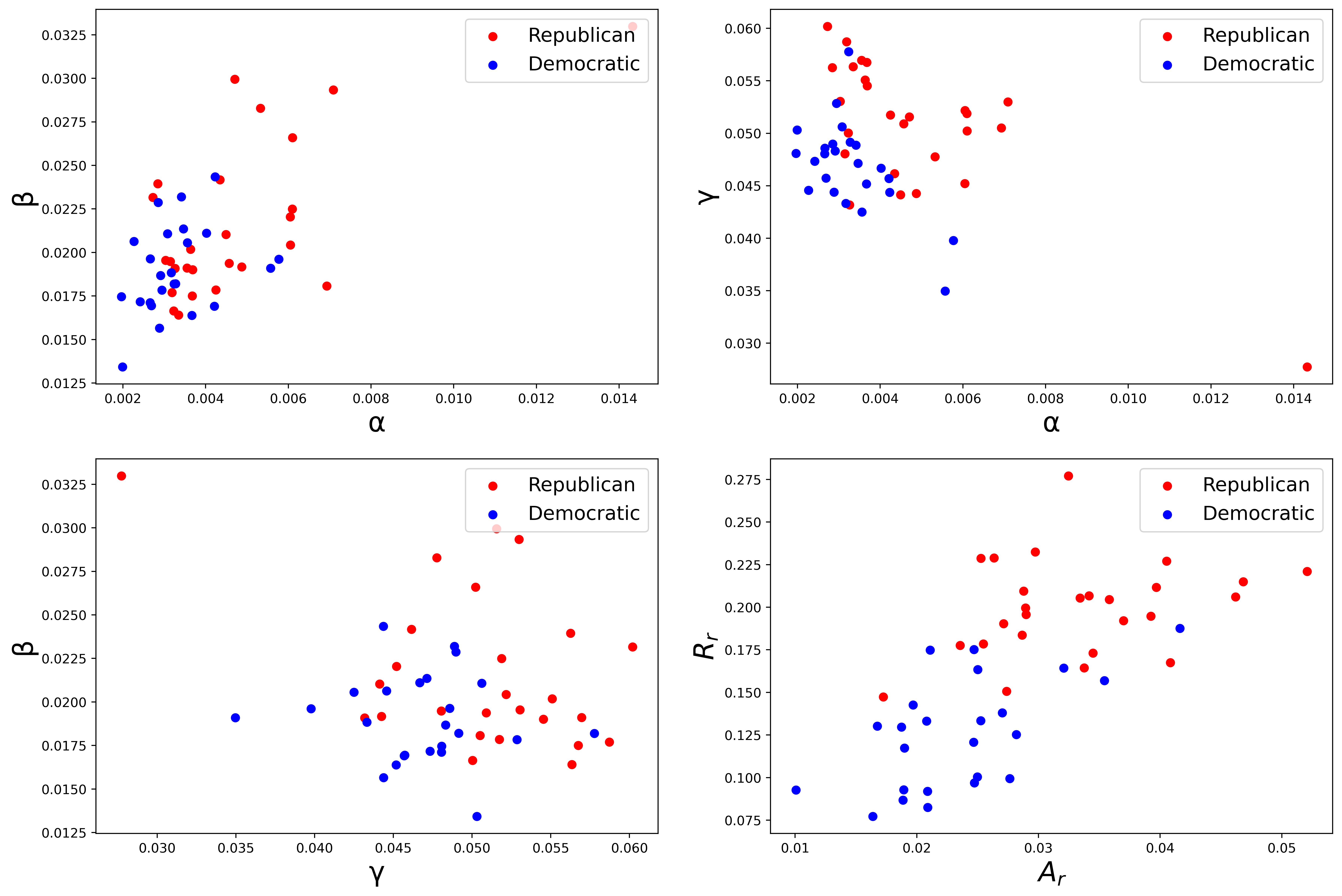}
    \caption{Scatter plots of optimised model parameters for different states based on uVAR Model~\eqref{equ: uVAR-model} (Each panel compares two parameters, with the x-axis representing one parameter and the y-axis representing another across all states. The points are colour-coded to indicate whether the datasets correspond to Republican or Democrat states, illustrating how the parameter pairs contribute to class separation).}
    \label{fig:VAR_scatter_plots}
\end{figure}

\begin{figure}
    \centering
    \includegraphics[scale=.4]{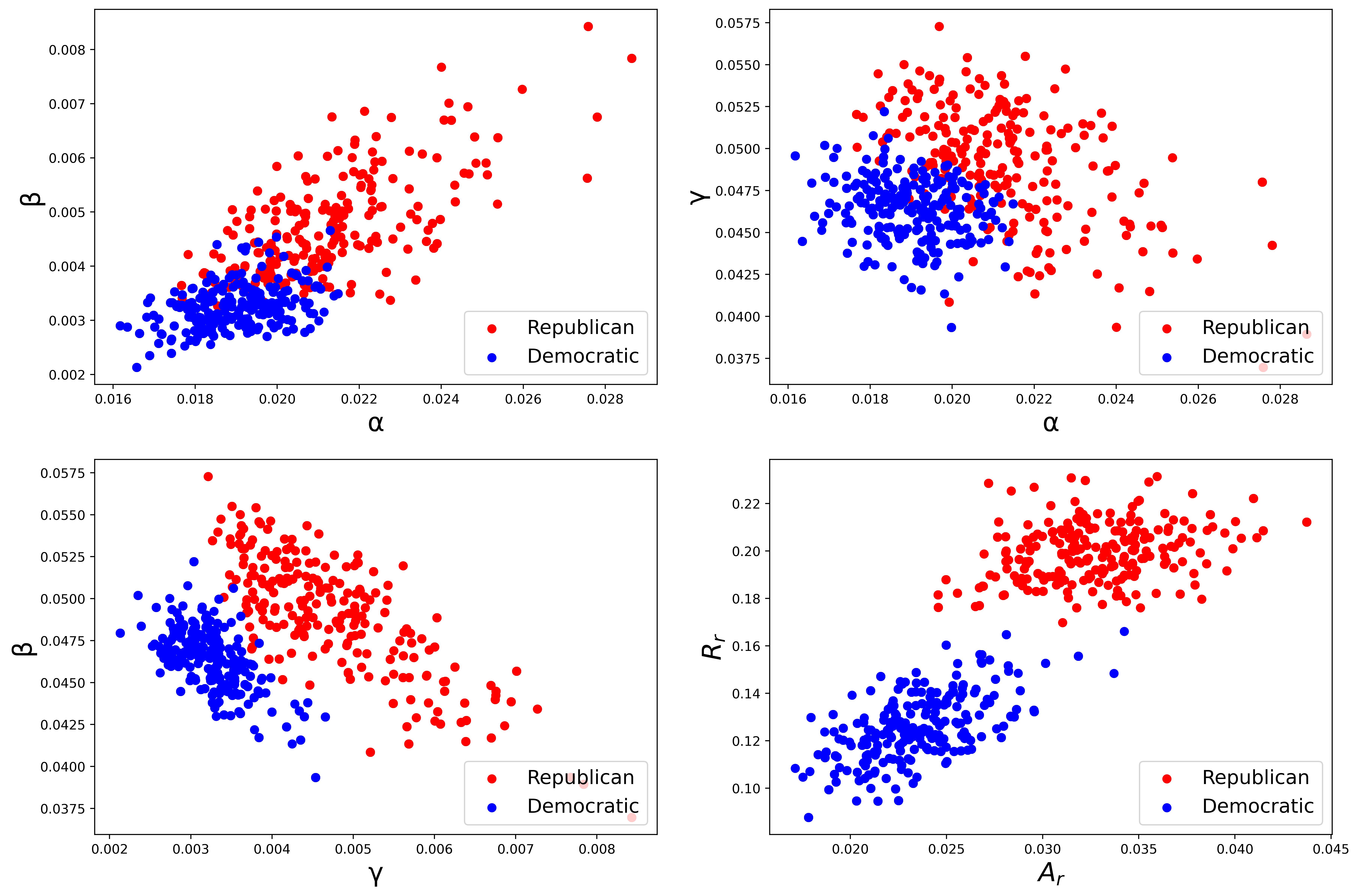}
    \caption{Scatter in each group (We randomly selected five states and aggregated them as a dataset, then we optimised our model and found parameters. We did it for each group 5000 times).}
    \label{fig:scatter_boots}
\end{figure}

\begin{figure}
    \centering
    \includegraphics[scale=0.3]{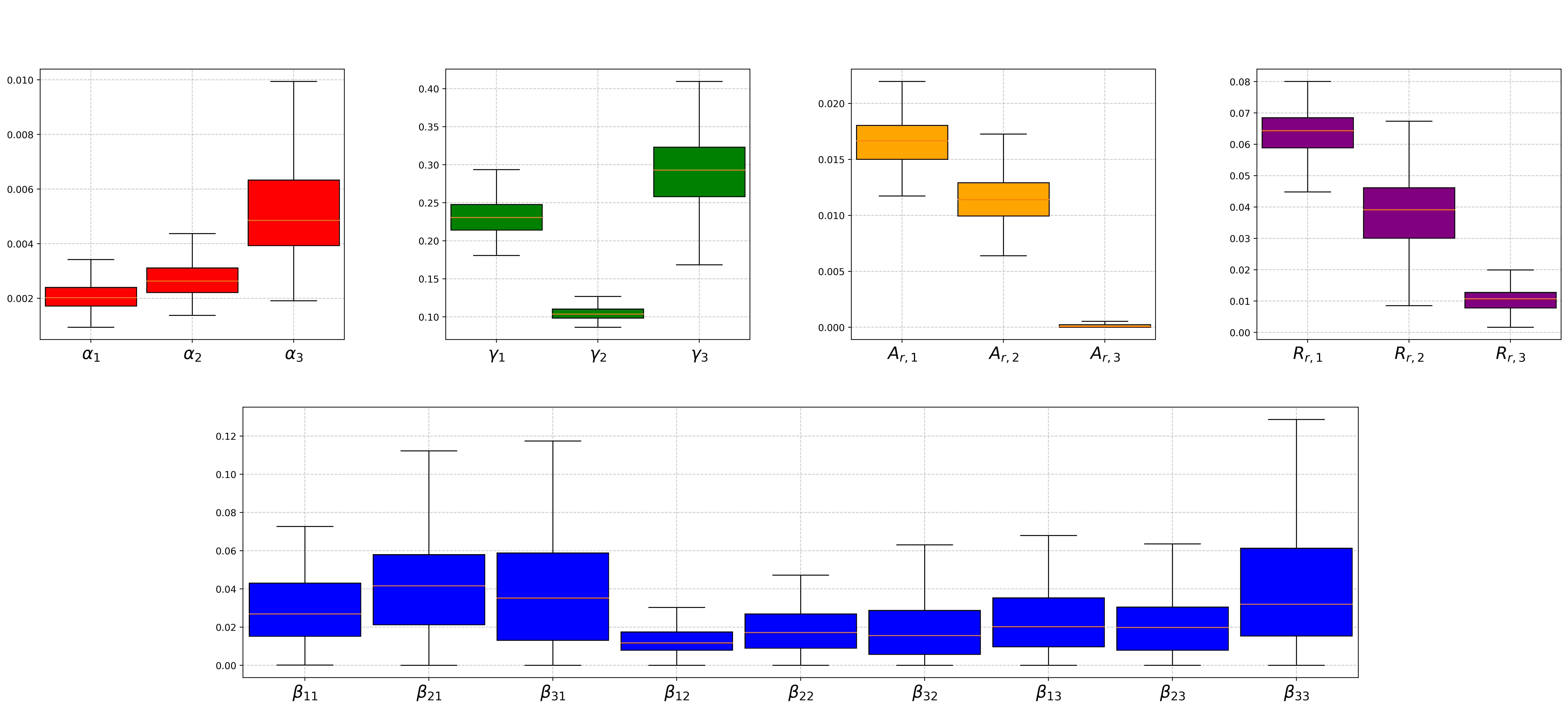}
    \caption{Boxplot of parameters in sVAR Model~\eqref{equ: sVAR-model} for age using Bootstrapping method (We randomly selected five states and combined them into a dataset. We then optimized our model and determined the parameters. This process was repeated 5000 times for each group).}
    \label{fig:age_Boostrap_boxplot}
\end{figure}

\begin{figure}
    \centering
    \includegraphics[scale=0.4]{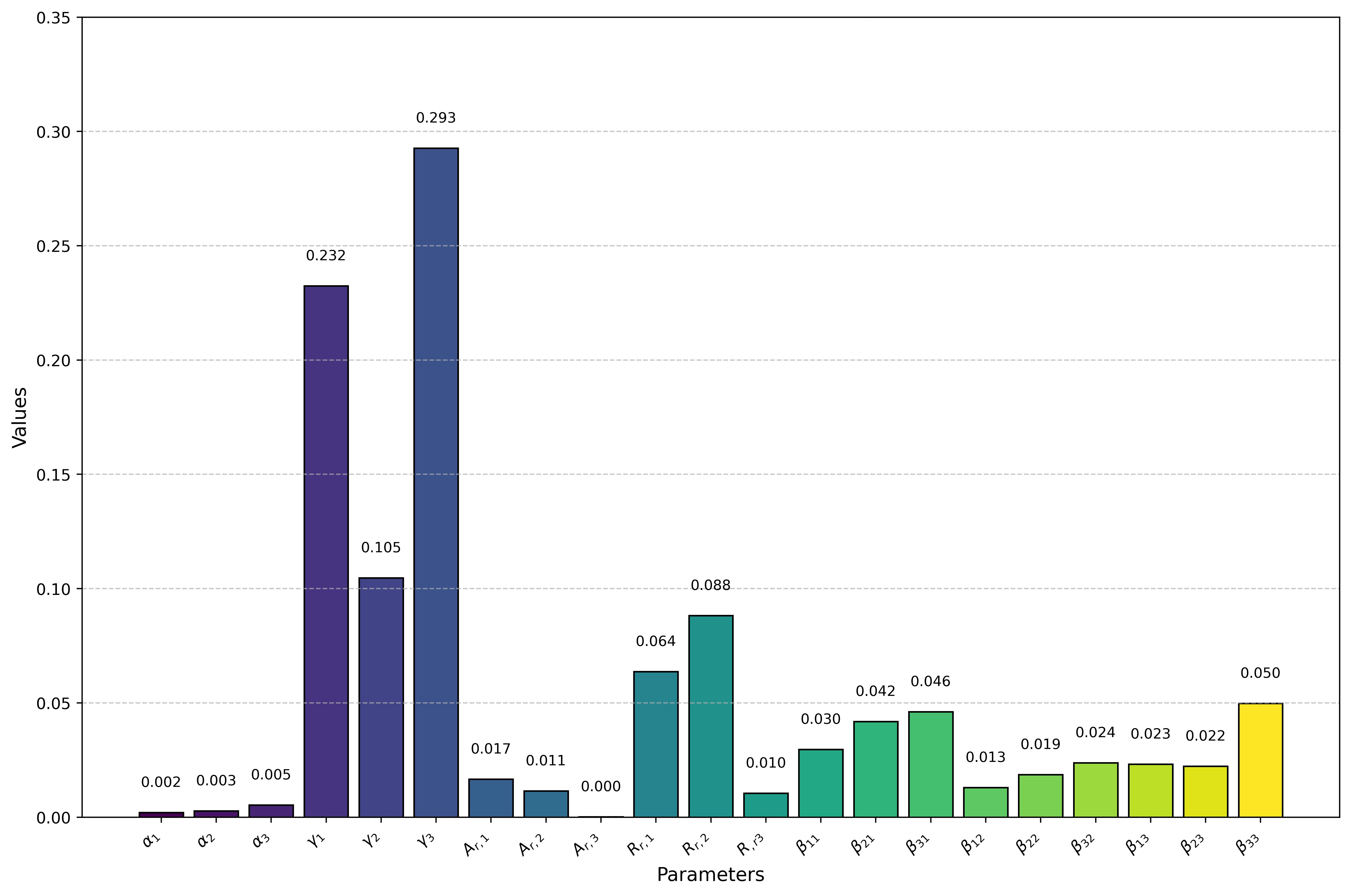}
    \caption{Average of parameters in sVAR Model~\eqref{equ: sVAR-model} for age using Bootstrapping method (We randomly selected five states and combined them into a dataset. We then optimized our model and determined the parameters. This process was repeated 5000 times for each group).}
    \label{fig:age_Boostrap_average}
\end{figure}

\begin{figure}
    \centering
    \includegraphics[scale=0.4]{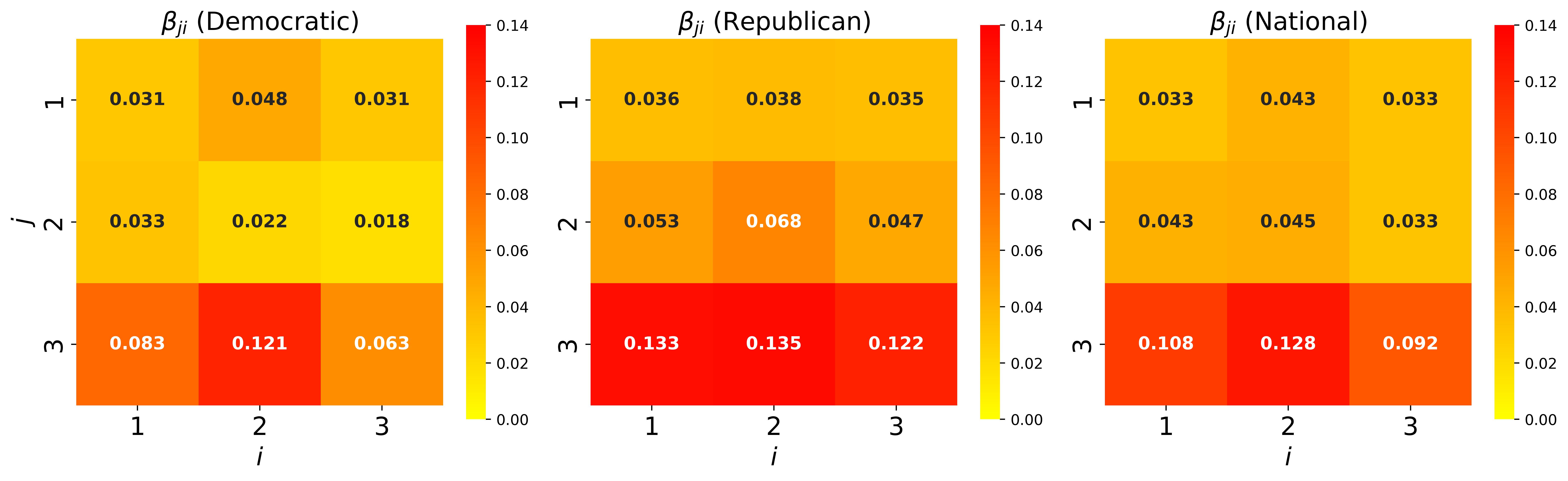}
    \caption{The matrix $[\beta_{ji}]$ of the average rates of influences of individuals in vaccinated or accepted class of age group $j$ on individuals in refused class of age group $i$ to change their opinion from refused to accept (The values of the parameters represent the average of the bootstraps distribution obtained by fitting sVAR Model~\eqref{equ: sVAR-model} to age-structured data).}
    \label{fig:age_matrix}
\end{figure}

\begin{figure}
    \centering
    \includegraphics[scale=0.3]{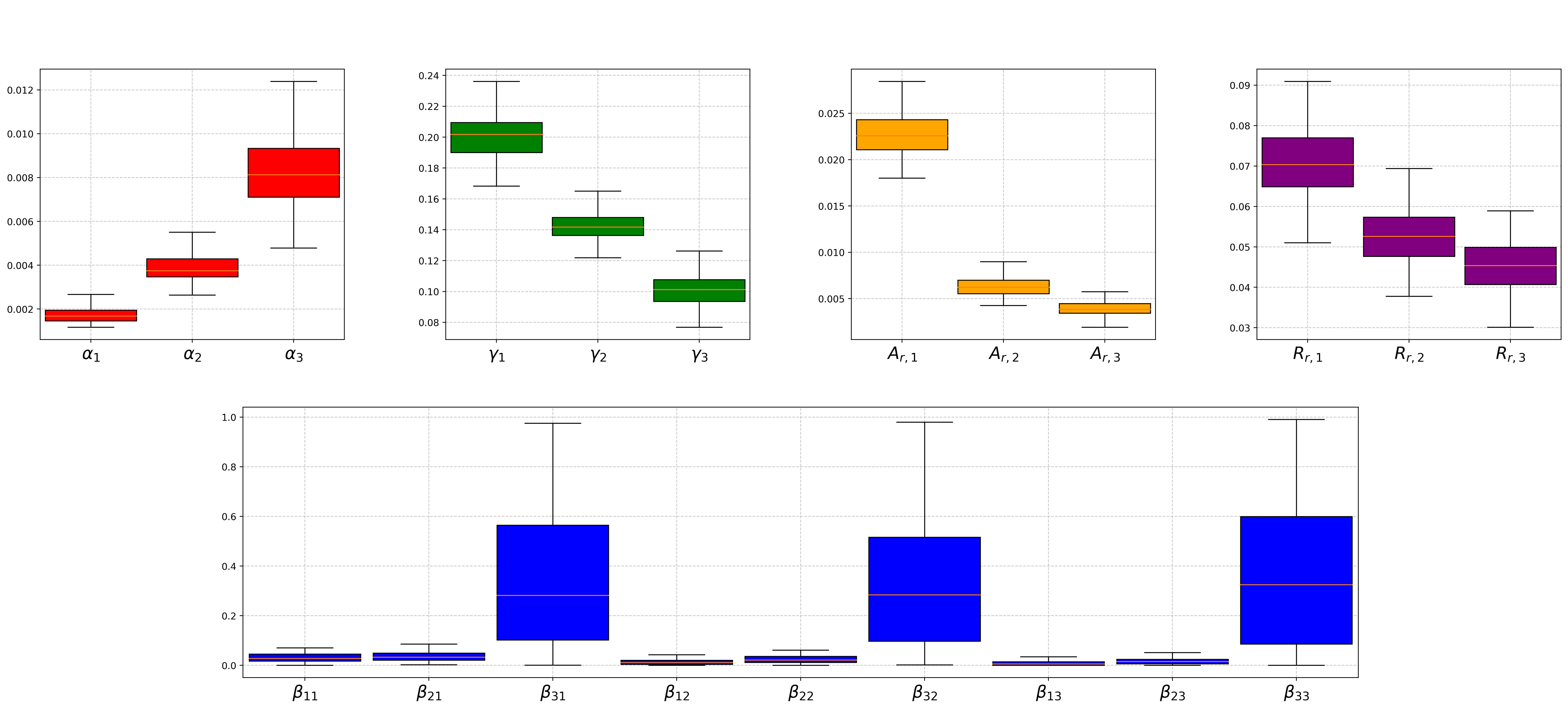}
    \caption{Boxplot of parameters in sVAR Model~\eqref{equ: sVAR-model} for education using Bootstrapping method (We randomly selected five states and combined them into a dataset. We then optimized our model and determined the parameters. This process was repeated 5000 times for each group).}
    \label{fig:edu_Boostrap_box}
\end{figure}

\begin{figure}
    \centering
    \includegraphics[scale=0.4]{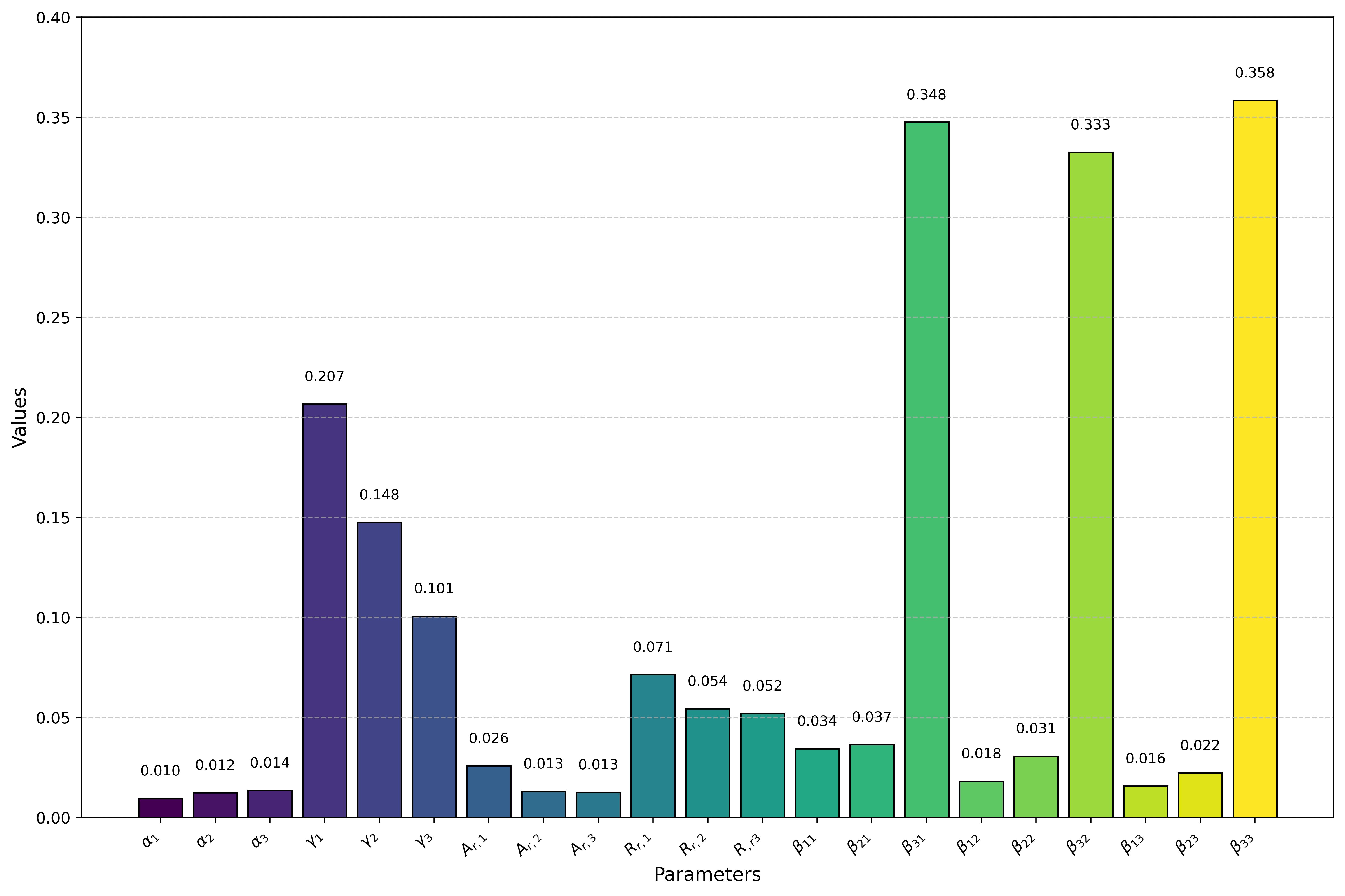}
    \caption{Average of parameters in sVAR Model~\eqref{equ: sVAR-model} for education using Bootstrapping method (We randomly selected five states and combined them into a dataset. We then optimized our model and determined the parameters. This process was repeated 5000 times for each group).}
    \label{fig:edu_Boostrap_average}
\end{figure}

\begin{figure}
    \centering
    \includegraphics[scale=0.4]{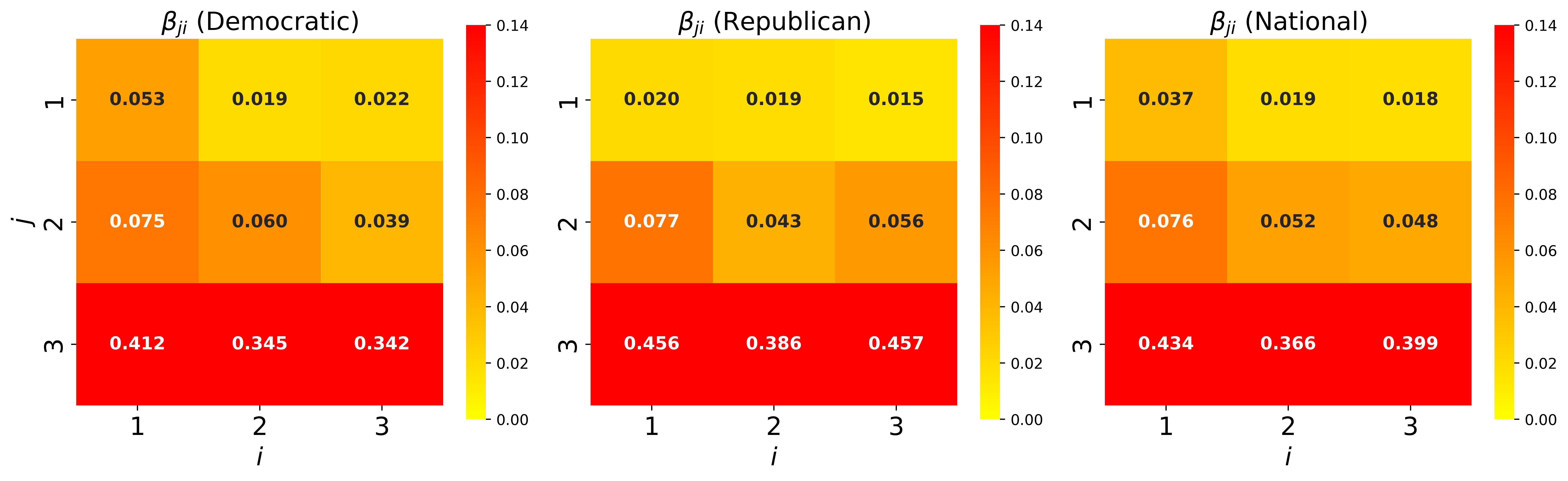}
    \caption{The matrix $[\beta_{ji}]$ of the average rates of influences of individuals in vaccinated or accepted class of education group $j$ on individuals in refused class of education group $i$ to change their opinion from refused to accept (The values of the parameters represent the average of the bootstraps distribution obtained by fitting sVAR Model~\eqref{equ: sVAR-model} to education-structured data).}
    \label{fig:edu_matrix}
\end{figure}
\end{document}